\documentclass[reprint]{revtex4-1}

\usepackage{bbold}
\usepackage{amssymb}
\usepackage{graphics}
\usepackage{graphicx}
\usepackage{epstopdf}
\usepackage{multirow}
\usepackage{amsmath}
\usepackage{amsthm}
\usepackage{amstext}
\usepackage{amscd}
\usepackage{epsfig}
\usepackage[mathscr]{eucal}
\usepackage{times}
\usepackage{srcltx}
\usepackage{shadow}
\usepackage{url}
\usepackage{color}
\usepackage[all]{xy}
\usepackage[normalem]{ulem}

\usepackage{enumitem}
\usepackage[most]{tcolorbox}

\usepackage{hyperref}

\begin{document}

\title{On the possibility of engineering social evolution in microfluidic environments}
\author{Gurdip Uppal, Dervis Can Vural}
\affiliation{Department of Physics, University of Notre Dame, Notre Dame, IN 46556}


\date{\today}

\begin{abstract}
Many species of microbes cooperate by producing public goods from which they collectively benefit. However, these populations are under the risk of being taken over by cheating mutants that do not contribute to the pool of public goods. Here we present theoretical findings that address how the social evolution of microbes can be manipulated by external perturbations, to inhibit or promote the fixation of cheaters. To control social evolution, we determine the effects of fluid-dynamical properties such as flow rate or boundary geometry. We also study the social evolutionary consequences of introducing beneficial or harmful chemicals at steady state and in a time dependent fashion. We show that by modulating the flow rate and by applying pulsed chemical signals, we can modulate the spatial structure and dynamics of the population, in a way that can select for more or less cooperative microbial populations. 
\end{abstract}

\maketitle

\onecolumngrid
\begin{tcolorbox}[width=\textwidth]
Significance: Controlling the social evolution of microbial populations within microfluidic environments holds immense promise across medical, environmental, and agro-industrial sectors. Experimental evolution within microfluidic systems remains conspicuously limited, and realistic models to guide our understanding are missing. Here, we explore a physics-based, first-principles model of microbial evolution in fluid dynamic environments and show different modes of social evolution in microbes as a function of different flow patterns, domain geometries and chemical composition. Based on these theoretical observations, we then explore the potential to tune physical parameters to deliberately enhance or suppress social cooperation of microbes.
\end{tcolorbox}
\twocolumngrid
\section*{Introduction}

Many microorganisms communicate and cooperate through diffusing secretions. Microbes gain strength in numbers and together they facilitate the decomposition of organic waste \cite{pan2012composting}, act as biofilters \cite{cohen2001biofiltration}, aid the digestion of food in our guts \cite{zhang2015impacts}, and harvest solar energy \cite{cogdell1999photosynthetic}. In medical, industrial and environmental applications it can be essential to maintain and promote the cooperative coexistence of microbes. For example, in the activated sludge in a water treatment facility, or in our guts, it is desirable for the microbial community to cooperate and stably coexist \cite{xia2018diversity,dore2015influence}. Of course, microbial cooperation is not always desirable. Cooperative aggregates of microbes more easily gain antibiotic resistance \cite{lai2009swarming,stewart2002mechanisms,cox2013intrinsic} and can cause the biofouling of marine vessels \cite{de2018marine,schultz2011economic}. Cancer cells cooperate by means of secreting vascularization factors, decaying the extracellular matrix and nearby normal cells, and manipulating the acidity of their microenvironment. Cancer cells can also cooperate to coerce normal fibroblasts to promote cancer growth, through use of diffusible factors to induce the stroma to help cancer cells \cite{joyce2009microenvironmental}. 

Controlling social evolution is therefore of considerable interest to combat infections, to utilize microbial functions in industrial and environmental applications, and to develop novel medical treatments. More generally, understanding what factors shape the social interaction structure of organisms can also offer insight into the origins of complex, multicellular life \cite{palkova2004multicellular}. 




Many biotic and abiotic factors influence the social evolution of microbes. Fluid flow has recently been shown to disrupt quorum communications in biofilms \cite{kim2016local,stoodley2016biofilms}. Microbes themselves have evolved mechanisms to control cheating, such as with policing, where cheaters are inflicted with a cost \cite{wechsler2019understanding}.
Fluid flow has been suggested to enhance group fragmentation and thereby promote sociality in microbes \cite{uppal2018} and also modulate microbial specialization and coexistence \cite{uppal2020evolution}. Disturbance has also been shown to influence cooperation in microbes, with cooperation peaking at intermediate disturbance \cite{brockhurst2007cooperation}. 

In cancer, where cells clonally reproduce, high relatedness and repeated interactions can lead to evolutionarily stable cooperative strategies in public good games \cite{archetti2019cooperation}. The exploitation and utilization of these evolutionary mechanisms can facilitate intentional control over the social evolution of cells for practical purposes \cite{cavaliere2017cooperation}. 

Many current strategies for designing and controlling microbial populations rely on targeting individual-level traits. For example, public goods such as prebiotics \cite{sheth2016manipulating} have been used to either promote the growth of certain microbes \cite{roberfroid2010prebiotic,bouhnik2004capacity} and toxins, such as antibiotics  have been used to inhibit the growth of others \cite{kohanski2010antibiotics,robinson2010antibiotic}. These methods are simpler to understand but fail to capture the complex interactions between members of the population. As a result, bacteria are selected for antibiotic resistance \cite{murray2018novel} and intra-tumor heterogeneity leading to the selection of drug resistant cells in cancers \cite{mcgranahan2015biological}. In contrast, exploiting and controlling the interaction structure of the population offers new ways to make use of or to combat microbial populations \cite{de2013bacterial,taylor2014antibiotic,morsky2022suppressing}. Understanding how to control the interaction structure can help utilize the full potential of microbial populations \cite{cavaliere2017cooperation,zhai2017microbial}.
 

Recent experiments show promise in understanding and controlling the evolution of social behavior of microorganisms. For example, it is possible to use quorum blocking drugs to blind bacteria of each others presence. In the particular case of \emph{Pseudomonas aeruginosa}, the bacteria stop producing a particular public good, which help them to extract iron from a host. Since the microbes resistant to the quorum blocker will continue to produce the costly molecule, they are taken over by their selfish counterparts affected by the drug. Eventually, the iron-deficient population is easily defeated by the hosts immune system  \cite{boyle2013exploiting}. Synthetic communities can also be engineered to enhance production of natural products by distributing long metabolic pathways among microbial consortia, even when a natural mutualism is not present \cite{zhou2015distributing}.

Experiments with a lung cancer model show promise in steering the evolutionary trajectory of cells by utilizing trade-offs where resistance to one drug leads to increased sensitivity to subsequent treatments \cite{acar2020exploiting}. Game theoretic studies of cancer models also propose promising methods in controlling cancer. One strategy called ``adaptive therapy'' proposes using a smaller dose of drugs, rather than maximally killing tumor cells, to encourage cell-cell competition and slow down the development of resistance \cite{gatenby2009adaptive,enriquez2016exploiting, gallaher2018spatial}. Preliminary clinical trials with this approach show promising results \cite{zhang2017integrating}. Another approach could be to engineer tumor cells by knocking out the genes coding for essential growth factors. These cells would then benefit from others that produce growth factors and lead to a tragedy of the commons, acting as a ``tumor within a tumor'' \cite{archetti2019cooperation}.



Many theoretical studies that investigate the evolution of the interaction structure of many microbes typically use effective game theoretic methods and fail to capture the significance of physical and spatial effects. These models typically assume well mixed populations, fixed group size, or effective phenomenological spatial structures \cite{szabo,Allen2013,nowak04,dcv2}. While there are a few models that take into account spatial proximity effects \cite{Medvinsky2002,Nadell2010,nadell2013,Dobay2014,Driscoll2010,wakano2009} as well as the decay and diffusion of public goods \cite{wakano2009,Dobay2014,hauert2008,menon2015public}, how flow and geometry effect the social evolution of a population remains mostly unexplored.

In this paper we propose multiple mechanisms that can be used to control social evolution through external abiotic control methods. Specifically, we investigate the influence of perturbing the system with toxins and goods at well defined time intervals; and using various channel geometries and domain shapes to induce types of flow patterns that can suppress or enhance the evolution of cooperation. 

We explicitly model the microbes and their interactions starting with first principles, taking into account the effects of fluid flow patterns, molecular diffusion and decay constants, and cell growth kinetics. The group structure observed in our model is emergent, allowing us to investigate the effects of flow, chemical perturbations, and geometry on the formation, fragmentation, and mixing of social microbial groups.

Using physics based first-principles simulations, we find that cooperating microbes self-aggregate into communities under certain physical conditions; and show that the spatial structure and social behaviors of these communities can then be modulated systematically by introducing various physical perturbations externally.

Our computational model can be characterized by the following assumptions, stated qualitatively: (1) each microbe secretes one (fitness enhancing) public good with some metabolic cost, and one (fitness reducing) waste byproduct with no cost.  (2) random mutations can vary the secretion rate of the public good, (3) microbes and their secreted compounds flow and spread according to the laws of fluid dynamics and diffusion, as dictated by the domain geometry.

We find that the physical effects of the environment strongly influences the social evolution of microbial species. Through evolutionary computer experiments and analytical formulas, we see the social behavior of microbes can be manipulated through (1) external chemical perturbations, (2) modulating the domain geometry and (3) the fluid flow profile. 

Steering the social evolution of microbial populations can give novel strategies to deal with microbial infections, biofouling, and to utilize biofilms for waste treatment. Understanding the factors contributing to sociality is also interesting from evolutionary biology and ecology stand points, as such factors illuminate how multicellularity and biological complexity originates and is stably sustained. 

\begin{figure}
\centering
\includegraphics[width=\linewidth]{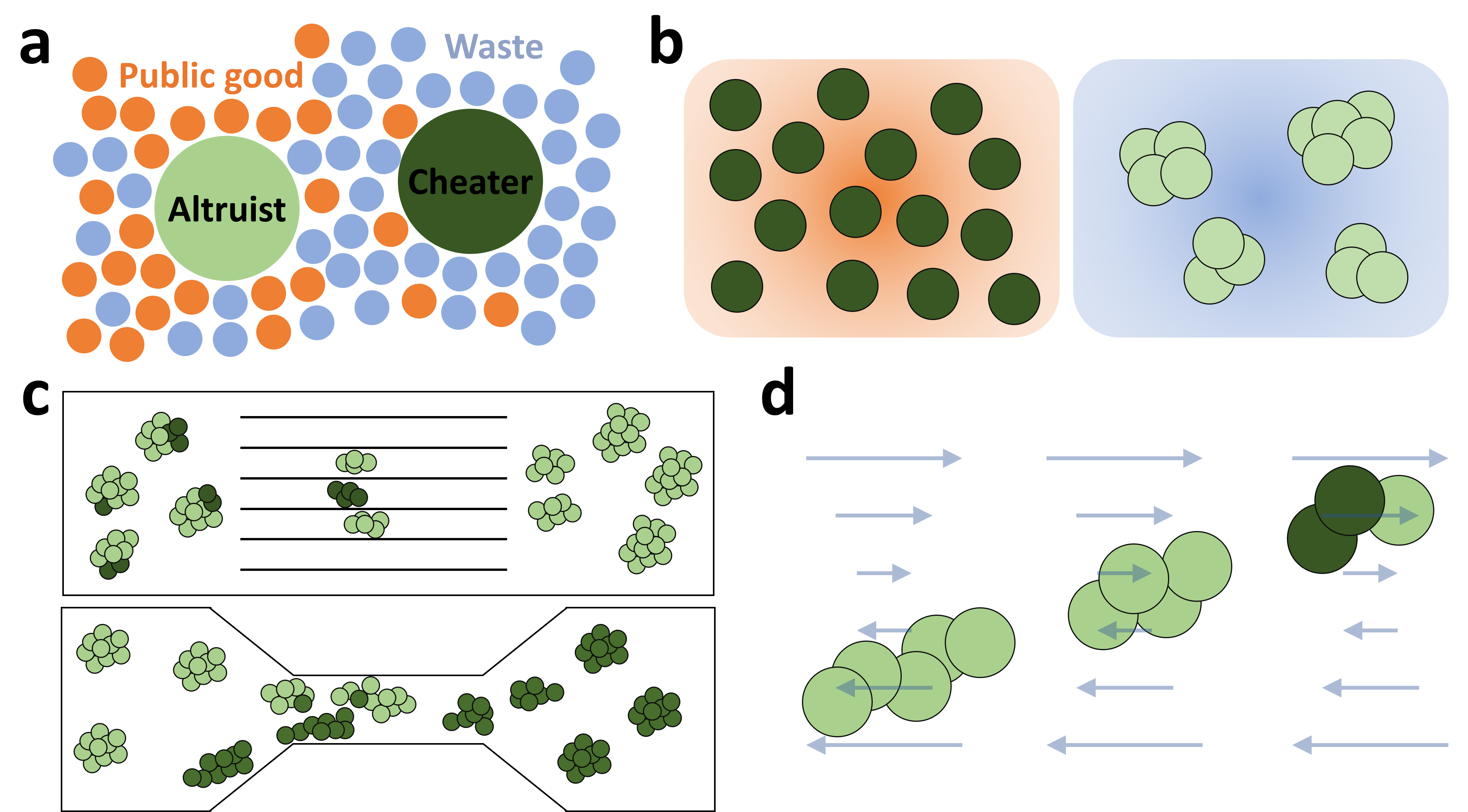}
\caption{{\bf Schematic of model and control strategies.} {\bf (a)} Microbes secrete two types of molecules into the environment. The first, a beneficial public good that promotes growth, and the second, a waste or harmful substance that hinders growth. Cheating microbes produce less or none of the former, while benefiting from public goods secreted by the cooperating population. We explore which strategies lead to more cooperative and less cooperative behavior as well as the localization of cooperation in space. {\bf (b)} Controlling the growth and evolution of microbes by externally introducing public good and toxin chemicals. Externally adding the public good or toxin alters group formation and the social evolution of microbes. {\bf (c)} Controlling evolution through geometry. By varying the geometry, we can ``filter out'' (above) or ``mix in'' (below) cheaters from cooperating groups. {\bf (d)} Control through fluid flow. A shearing flow fragments groups and limits the spread of cheaters.}
\label{fig:model}
\end{figure}

\section*{Model}
We study an evolving, spatial population where social behaviour can be modulated by introducing beneficial or deleterious drugs, changes in the domain geometry, and structure of fluid flow. Our work consists of discrete, stochastic agent-based simulations as well as closely related continuous deterministic formulas.

In our model, the social microbes secrete two types of diffusive molecules that influence each other's fitness (Fig. \ref{fig:model}): One is a public good, whose local concentration is denoted by $c_1(x,t)$, that increases the growth rate of nearby microbes, including the producer. The second, $c_2(x,t)$, is a waste compound / toxin that curbs the growth of nearby microbes, including the producer. We assume that the public good incurs a metabolic cost to a producer, whereas the waste compound does not. Furthermore, we assume that the secretion rate $s_1$ of the public good can mutate, whereas that of the waste molecule, $s_2$ does not. These assumptions are realistic, as the release of a waste molecule typically results from a core metabolic process, imposing no extra metabolic cost and remaining immutable. In contrast, the production of public goods typically incurs a cost, and microbes may undergo mutations affecting their ability to produce them.

We also include control functions that allow us to externally input drugs that can similarly enhance or curb the growth of cells. We explore the evolutionary outcomes of treating the system with such drugs steadily as well as with time-dependent pulses.

We simulate the cells as discrete particles subject to stochastic physical and evolutionary forces, and the secreted compounds and drugs as continuous fields (see Methods). In contrast, our analytical expressions are derived from the following deterministic, continuous differential equations,

\begin{align}
&\dot{n}  =  d_b \nabla^2 n - \mathbf{v} \cdot \mathbf{\nabla} n + n f(c_1, c_2, s_1)  + \sigma \frac{\partial^2 n}{\partial s_1{}^2} , \label{eq:microbes} \\
&\dot{c}_1 = d_1 \nabla^2 c_1 - \mathbf{v} \cdot \mathbf{\nabla} c_1 + n s_1 - \lambda_1 c_1 + \mu_1 (t), \label{eq:goods} \\
&\dot{c}_2 = d_2 \nabla^2 c_2 - \mathbf{v} \cdot \mathbf{\nabla} c_2 + n s_2 - \lambda_2 c_2 + \mu_2 (t), 
\label{eq:waste}
\end{align}
where the growth rate $f(c_1,c_2,s_1)$ governed by the local concentrations $c_1(x,t)$ and $c_2(x,t)$, is given by

\begin{align}
&f(c_1, c_2, s_1) = \alpha_1 \frac{  c_1}{c_1 + k_1} - \alpha_2 \frac{ c_2}{c_2 + k_2} - \beta_1 s_1 .
\label{eq:fitness}
\end{align}

Here $n$ is a shorthand for $n(x,t,s_1)$, the number density of bacteria that produce the public good at a rate of $s_1$. These bacteria pay a metabolic cost of $\beta_1 s_1$ per unit time, which accordingly decreases their growth rate. The production rate of waste $s_2$ is assumed constant for all bacteria and has no cost to its producer. Waste limits the number of bacteria that a region can carry, effectively acting as a local carrying capacity. Without the waste term, the population may undergo unrealistic, unlimited growth. Bacteria producing the public good at a rate $s_1$ produce more of the same at a rate given by equation \ref{eq:fitness}. However, the production rate $s_1$ can change due to mutations. This is described by the last term of equation \ref{eq:microbes}. In the continuous equations, mutations can be thought of as a diffusion in $s_1$ space, whereas in our simulations, we treat mutations as a corresponding discrete random walk.

In all three equations the first two terms describe diffusion and flow; the third and fourth terms of equations \ref{eq:goods} and \ref{eq:waste} describe the production and decay of chemicals. The last terms in equations \ref{eq:goods} and \ref{eq:waste} represent our external control. The functions $\mu_1 (t)$ and $\mu_2 (t)$ allow us to introduce chemicals independently in either a constant or time dependent fashion. We focus only on controls homogeneous in space. 

The first two terms in equation \ref{eq:fitness} describe the effect of the secreted compounds on the fitness of bacteria. This saturating form is experimentally established \cite{Monod1949}, and well understood and commonly used in population dynamics models \citep{allen2018bacterial}. The crucial third term in equation \ref{eq:fitness} describes the cost of producing the public good. The cost grows linearly with the public good secretion rate.

We study the evolutionary outcomes of this model both theoretically and through simulations. In the computational analogue of these equations, we treat bacteria and their mutations as discrete entities (see Methods). 

\section*{Results}
\subsection*{Controlling social evolution through drugs}
If the diffusion length of the public good is smaller than that of the waste compound, we find that microbes cluster together to form cooperative groups. These clusters are key for the evolution of cooperation. When the diffusion length of the public good is larger than that of the waste compound, clusters do not form, and in this case, cheating mutants (those who do not secrete the public good) spread throughout the entire population and take over, leading to a tragedy of the commons scenario, and the population goes extinct. The formation of these microbial clusters can be understood analytically in terms of the ``Turing mechanism'', which we studied in an earlier paper \cite{uppal2018} (also cf. Appendix \ref{app:turing}). 

\begin{figure*}
\centering
\includegraphics[width=0.95\textwidth]{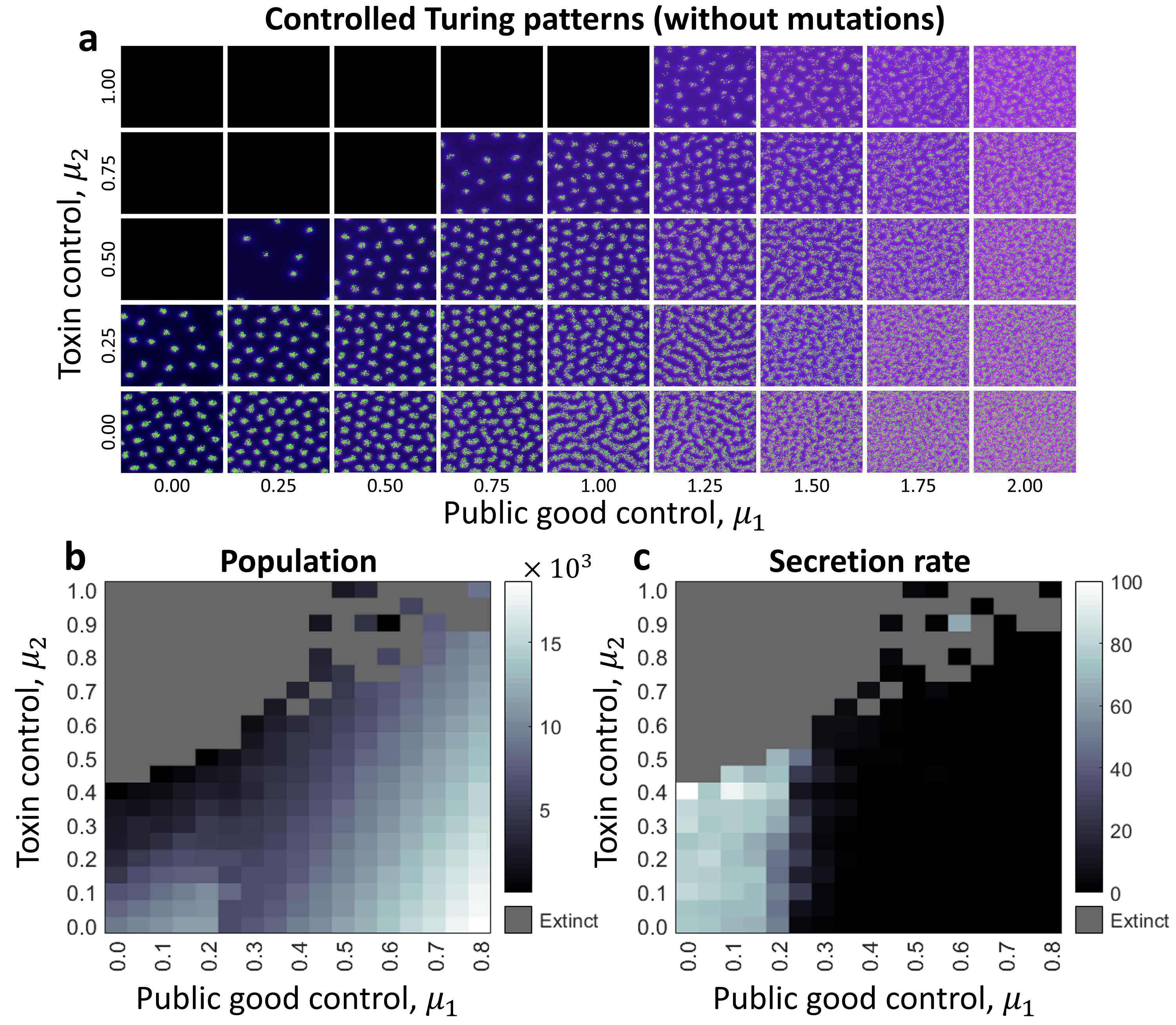}
\caption{{\bf Control via constant chemical injection.} {\bf a,} Control of Turing patterns with added chemicals. Adding more of the toxin to the system decreases the population density. As we add more public good, the system becomes denser, more homogeneous, and no longer shows Turing patterns. Cheaters are then able to spread throughout the population and take over. By adding the public good and toxin together, we can maintain the group structure and control the group size, within some limits. {\bf b,} Average population for mutating population with chemical controls constant in time. Adding more public good allows cheaters to be stable without the presence of cooperators. The population in general grows with added public good and declines with added toxin. The population goes extinct in the top left region where the added concentration of the toxin is too high and the public good is low. {\bf c,} Average secretion rate for mutating population with chemical controls constant in time. Adding more public good to the system selects for cheaters. As we add more public good, the system no longer shows Turing patterns and becomes homogeneous. Cheaters are then able to spread throughout the population and take over. We see this occur when the added public good is greater than around $0.2$. The constant public good allows cheaters to be stable. Adding the toxin forces cheaters to secrete more public good to maintain fitness, but lowers the overall population density. This makes the system more susceptible to extinction. The top-left regions where panels {\bf b,c} are grey correspond to the population being extinct due to too much added toxin. The initial public good secretion rate was set to $s_1 = 100$, the mutation rate for panels {\bf b,c} was set to $\sigma = 1 \times 10^{-7}$, and flow rate was set to zero. All other parameters are as given in Table \ref{tab:parameters}.  }
\label{fig:constantcontrol}
\end{figure*}

We first study the effects of adding the public good and waste terms externally into the system in a controlled manner. We see how adding these chemicals modulates the geometry of emergent clusters and how this affects the social evolution of the microbes.

{\bf Constant application of drugs.} We first explore adding drugs (specifically the public good or toxin) homogeneously in space and constant in time. We find by varying the amount of added public good and/or toxin, we can drastically alter the structure of clusters (Fig. \ref{fig:constantcontrol}a). Adding enough public good and/or toxin will cause the spot pattern observed in the native state to transition to a striped state or a homogeneous state (Fig. \ref{fig:constantcontrol}a). In the striped and homogeneous states, the cooperative structures are no longer spatially isolated. These systems are now more susceptible to the spread of deleterious mutations that do not secrete as much or any of the public good. The added chemicals also affect the population density. Adding too much of the toxin will cause the system to go extinct. Adding more public good will increase the population until it saturates.

When we introduce mutations, we see that above a critical amount of externally added public good, cheaters take over the system (Fig. \ref{fig:constantcontrol}b,c and Supplementary video \ref{itm:const_pg}). This is because adding the external chemicals suppress the Turing patterning and transforms the system into an homogeneous state, where cheaters are able to spread throughout the full population. The effect of adding the public good also allows cheaters to sustain without the need for cooperators (defecting cheaters cannot otherwise survive without cooperators in our model). In addition, more public good also allows for a larger stable population (Fig. \ref{fig:constantcontrol}b). 

We therefore see three regions of sociality when adding controls. For low amounts of added public good and moderate amounts of added toxin, sociality prevails and microbes have a large secretion rate (bottom left region of Fig. \ref{fig:constantcontrol}c). When the added toxin is too large and public good is lower, the population goes extinct (top left region of Fig. \ref{fig:constantcontrol}b,c). When the added public good is too large and the toxin levels are lower, the population gets taken over by cheaters that are stable and grow large in number (bottom right region of Fig. \ref{fig:constantcontrol}b,c).

Adding the toxin externally can pressure the system to remain more cooperative (c.f Supplementary video \ref{itm:const_toxin}). With the added toxin, microbes must secrete more public good to compensate for the loss in fitness. However, even though microbes are more cooperative, the overall population of the system is lowered, and the system becomes more fragile, as seen in Fig. \ref{fig:constantcontrol}b,c for large toxin control values. Therefore, in order to use chemicals to increase sociality and population, we need to do more than just introduce them constantly.



\begin{figure*}
\centering
\includegraphics[width=0.98\textwidth]{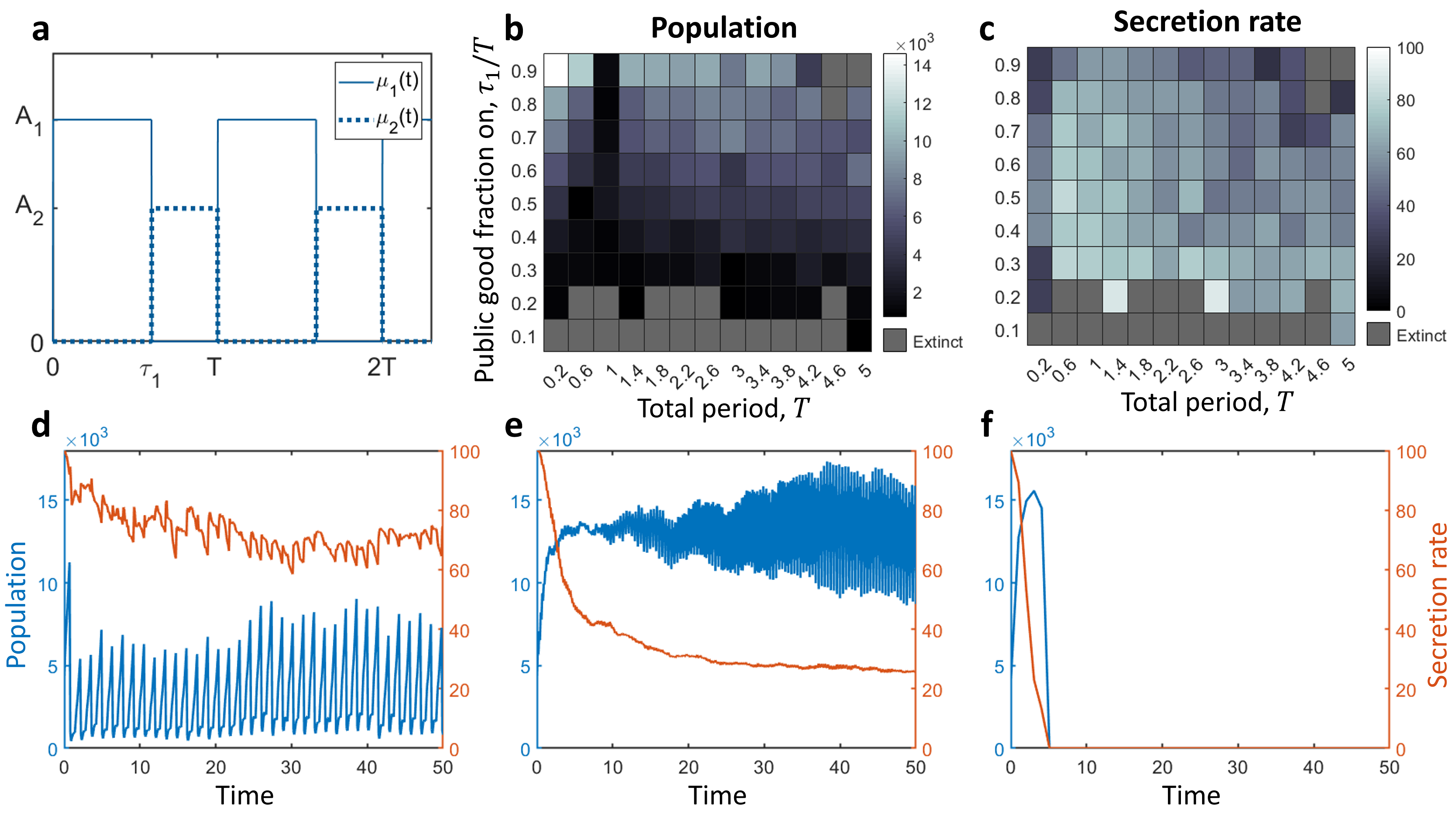}
\caption{{\bf Controlling social evolution via pulsed chemical injection.} {\bf a,} Schematic of square pulse control function. Chemicals are added periodically, starting with the public good with an amplitude of $A_1$ for a duration of $\tau_1$, followed by the toxin with an amplitude of $A_2$ added for a duration of $\tau_2$, and then repeated every period, $T = \tau_1 + \tau_2$. {\bf b,} Average population for mutating population with chemical controls varying in time. We plot the population versus the total pulse period along the x-axis and the fraction of time corresponding to the public good pulse along the y-axis. With the larger mutation rate, a pulse comprised solely of the toxin or even a majority of toxin goes extinct. Groups do not fragment quick enough and cheaters take over, leading to extinction. On the other end, a pulse solely comprised of public good drives the system to be comprised solely of cheaters with a large population. If the total period is too large, we also see the system goes extinct as cheaters take over. The slow period does not help groups fragment quicker than they usually would. {\bf c,} Average secretion rate for mutating population with chemical controls varying in time. Cooperation is mainly favored in the region where the total period is less than the native group reproduction rate (around $\omega = 1.5$) and where public goods comprise around 30-60\% of the total pulse.  {\bf d,} Time evolution in system with pulses leading to stable cooperation. The enhanced group fragmentation allows cooperators to persist at higher average secretion rates. {\bf e,} Time evolution in system with pulses leading to a stable population of cheaters. Here, a short total period predominately comprised of public good drives system to cheating state. {\bf f,} Time evolution in system with pulses leading to extinction. An initial long pulse of public good drives the population to be comprised of cheaters. A small subsequent toxin pulse then easily drives the population to extinction. The initial public good secretion rate was set to $s_1 = 100$, mutation rate was set to $\sigma = 1.5 \times 10^{-7}$, and flow rate set to zero. The amplitude for public good pulses was set to $A_1 = 0.5$ and for toxin pulses $A_2 = 0.3$. Simulations were run in a domain of size $40 \times 40$ for a duration of $T = 50$. All other parameters are as given in Table \ref{tab:parameters}. }
\label{fig:pulsecontrol}
\end{figure*}

Controlling both population and secretion rate can be important for practical purposes. Controlling the sociality can allow for new treatments against bacterial infections \cite{boyle2013exploiting}. In addition to controlling the sociality, we would also want to control the abundance  of microbes. For example, to help fight off a bacterial infection, one can select the control that gives the minimum abundance that is all cheaters (the region where secretion rate goes to zero and population is finite). 
We leave the control on long enough so that cheaters have taken over the system, thereby driving the population to be less cooperative. The microbes would then be in a more fragile state when the control is turned off and can be annihilated by the hosts immune system. We see this occur when adding the chemicals in a time-dependent scheme (Fig. \ref{fig:pulsecontrol}f).



{\bf Pulsed chemicals.} We can also promote cooperation by increasing the group fragmentation rate via injecting compounds periodically into the system. Fragmentation of groups is the key to cooperation since it limits the mobility of cheaters. As we have shown above, adding the chemicals constantly in time either selects for cheaters, or selects for cooperators, but with a low population. 

To promote social evolution, we explore injecting chemicals in a periodic fashion to induce group fragmentation. Our protocol is as follows: we first uniformly apply a pulse of toxin. As a result, the groups fragment into smaller pieces as weak spots die off. We then immediately apply the public good to cause each of these individual fragments to recover and grow into new, stable groups. We then repeat the process to fragment, and then grow the fragments again. 

Quantitatively, we model each chemical pulse as,
\begin{align}
    &\mu_1 (t) = \begin{cases}
        A_1 & t \pmod{T} < \tau_1 \\
        0 & \text{Otherwise} 
    \end{cases} \\
    &\mu_2 (t) = \begin{cases}
        A_2 & t - \tau_1 \pmod{T} < \tau_2 \\
        0 & \text{Otherwise} 
    \end{cases} .
\end{align}
Here $A_i$ is the amplitude of the pulse for chemical $i$, $\tau_i$ is the duration for which the pulse for chemical $i$ is on and $T = \tau_1 + \tau_2$ is the full period for the protocol. Only one chemical is on at any time, starting with the public good at time $t = 0$. A schematic of the protocol is shown in Fig. \ref{fig:pulsecontrol}a.

When adding chemicals in this periodic fashion, we find we can have stable cooperation with a sustained population (Fig. \ref{fig:pulsecontrol}d). The mechanism at work is that the toxin pulses kill weaker areas with cheaters and fragment large groups into smaller ones. Subsequent public good pulses then allow the injured fragments to grow and recover. The next toxin pulse is then added before cheaters emerge and take over (Supplementary video \ref{itm:pulses}). This induced fragmentation is quicker than the natural fragmentation that occurs without our intervention. The chemical pulses therefore allow us to take an otherwise unstable anti-social system and turn it into a stable social one (Fig. \ref{fig:pulsecontrol}b,c). 

We find the average population and secretion rate varies with the total period of the protocol and duration of the public good pulse (Fig. \ref{fig:pulsecontrol}b,c). If the public good pulse duration is too small, groups do not fragment quick enough and cheaters take over, leading to extinction. In contrast, a pulse predominantly composed of the public good leads to stable cheater dominance with a high population. Also, if the total period of the protocol is much larger than the native group fragmentation rate, cheaters prevail as groups fail to fragment faster. We therefore find we can shift the population to different regimes of population and secretion rate by varying the pulse durations. The ideal pulse duration for a desired outcome will in general depend on the native group fragmentation rate -- as determined by growth rates, diffusion constants, and decay rates -- as well as on the mutation rate. We maintained constant amplitudes for public good and toxin pulses, promoting increased group fragmentation without causing extinction.


We see (Fig. \ref{fig:pulsecontrol}d) that the pulsed protocol allows us to maintain a tighter control on the social behavior of microbes. The average secretion rate stays relatively constant around the initial value. However, the abundance is not as well controlled with the pulsed control. Here the microbial abundance oscillates significantly due to alternating public good and waste pulses. On average however, the abundance is larger than our other sociality-enhancing control scheme where only toxins were added constant in time. Here, the additional presence of the public good boosts the population and facilitates group recovery from toxin pulses.

The mechanism offered here resembles the results of \cite{brockhurst2007cooperation}, where disturbance was shown to be a possible mechanism for the evolution of sociality. Here we exploit this mechanism to propose a way to engineer social behavior in microbes.

When the public good duration constitutes a larger fraction of the period, cheaters are able to take over (Fig. \ref{fig:pulsecontrol}e). If the public good is on for long enough before turning off, cheaters completely take over the system and grow large in number. Turing off the public goods then drives the population to extinction  (Fig. \ref{fig:pulsecontrol}f). This is similar to the concept of evolutionary steering studied for cancer cells in \cite{acar2020exploiting}. In general, the evolutionary adaptation of organisms to one type of molecule may come with an increased sensitivity to a complementary compound \cite{acar2020exploiting,zhao2016modeling}.

\subsection*{Controlling social evolution via channel / domain geometry}

We can also control the sociality of a microbial population by modifying the shape of the flow channel they inhabit. To achieve this, we introduce microchannels or boundary walls into a rectangular space to either (1) split our initial population of bacteria into multiple groups that are isolated from each other, or (2) to mix multiple groups together. Thus, by altering the geometry, we can make it easier or more difficult for cheating mutants to spread and thereby supress social behavior.

\begin{figure*}
\centering
\includegraphics[width=0.98\textwidth]{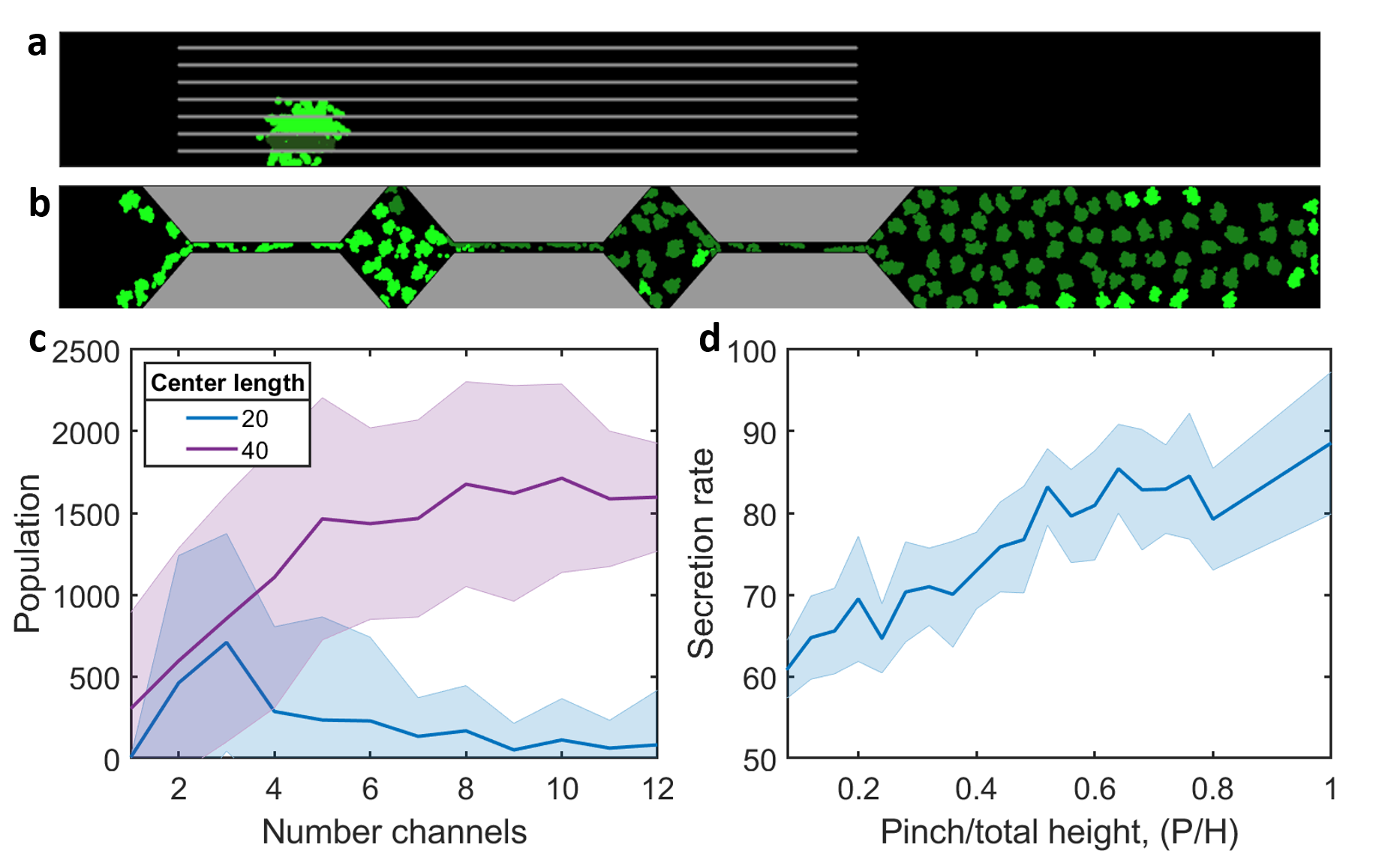}
\caption{{\bf Control via geometry.} {\bf a,b,} Simulation snapshots of using geometry to control sociality. {\bf (a)} Simulation snapshot of filter in action. A cooperating group enters the channeled portion of the filter and is split into segments. Segments with cheaters present get taken over and die out, whereas sections with cooperators pass through to reach the end of the filter. {\bf (b)} Snapshot of simulation with mixer geometry. Less social microbes (darker green) meet more social groups (light green) in a narrow channel and take over the more social groups. The groups that exit the narrow channel are then less social on average compared to geometries with wider channels. {\bf (c)} Final population for various filter channel widths. We see as we increase the number of channels (smaller channel widths), the cheaters are less likely to take over the population. By adding channels within the flow, we are able to fragment the group physically, thus making the system more robust against mutations. {\bf (d) } Final population and secretion rate for various central channel widths. For narrower channel widths, groups are more likely to come into contact and mix, resulting in a lower average secretion rate for the population. Parameter values for the filter, for public good diffusion was set to $d_1 = 5$, waste diffusion $d_2 = 500$, public good decay $\lambda_1 = 50$, waste decay $\lambda_2 = 3$, secretion rates $s_1 = s_2 = 25$, microbe diffusion $d_b = 0.2$, and fitness constants $\alpha_1 = 7$, $\alpha_2 = 8$, $k_1 = 0.01$, $k_2 = 0.1$, $\beta = 0.08$. Flow velocity was given by solving the Stokes equations with an inlet velocity of $v = 2$. For the mixer,  public good diffusion was set to $d_1 = 5$, waste diffusion $d_2 = 20$, public good decay $\lambda_1 = 50$, waste decay $\lambda_2 = 18$, secretion rates $s_1 = s_2 = 100$, microbe diffusion $d_b = 0.4$, and fitness constants $\alpha_1 = 60$, $\alpha_2 = 80$, $k_1 = 0.01$, $k_2 = 0.1$, $\beta = 0.2$.  Flow velocity was given by solving the Stokes equations with an inlet velocity of $v = 1$. Shaded regions correspond to one standard deviation from the mean, represented by solid lines. Averages were taken over 20 runs for each point.}
\label{fig:geometrycontrol}
\end{figure*}

{\bf Filtering with microchannels.} We first explore introducing micro channels that would split a group into multiple groups. Each separate section is then allowed to grow or die according to its local population. Channels in which there are cheating bacteria present die out before reaching the end of the filter, and channels without mutants make it through to the end. The channels therefore ``filter out'' the cheaters, leaving us with just cooperating bacteria at the end (Fig. \ref{fig:geometrycontrol}a). 

To simplify our analysis, we now place mutations in manually and explore the effects of geometry with no additional mutations. We start each simulation with one random group placed to the left of the filter (Fig. \ref{fig:geometrycontrol}a) and seeded with one cheater with $s_1 = 0$. We compare the social evolution of a group of microbes in geometries with varying number of microchannels in Fig. \ref{fig:geometrycontrol}c. We vary the geometry as follows: we fix the total height of the domain to be $H = 8$. We then add equally spaced walls to split the central region into microchannels. The walls are of thickness $W = 0.1$. The channel widths are then given as $C = (H - (N-1)W)/N$ for $N$ channels. Channels are added after a length of 7. We study channel lengths of 20 and 40. The region after the channels has a length of 50. We run the simulation for a time $T = 30$ and record the final population. Simulations where cheaters take over the whole group lead to extinct populations. We see that as we increase the number of channels, the cheaters are less likely to take over the population. As the filter splits apart the groups, channels with cheaters present die off, allowing social microbes to persist in isolated channels.

We also investigate the effects of varying channel length (Fig. \ref{fig:geometrycontrol}c). If the channels are too short, cheaters do not die out in time and are able to rejoin with cooperators at the end of the filter. The required length of the channels will in general depend on the invasion fitness of cheaters, the local population in a channel, and the flow rate. Channels need to be sufficiently long so cheaters take over their local population. Results can also be realized with shorter channel lengths if the flow rate is sufficiently low such that groups do not reach the end of the channel before all cheaters have died out.

We see that microchannels that happen to have cheaters, get dominated by these and die off as they pass through. Meanwhile, ``lucky'' channels that happen to have no cheaters, pass through safely. In contrast, if there were no channel filter, the cheaters would freely spread, and ultimately destroy the entire population. This observation aligns with Simpson's paradox \cite{chuang2009simpson, penn2012can}, where individual groups may decrease in sociality, but the population as a whole becomes more social.

In principle, by using such filters in select locations, one can also localize, or create spatial patterns of cooperation. 



{\bf Mixing.} In order to reduce the sociality of the population, we shape the flow domain to act as a mixer. Specifically, we add barrier walls to funnel microbes through a single narrow channel, thus forcing multiple groups to combine and mix together (Fig. \ref{fig:geometrycontrol}c). Groups that contain microbes with lower secretion rate will then seed groups with larger secretion rates. The cheating microbes will then dominate whichever group they find themselves in. 

Again, to simplify analysis, we start our simulation with one cheater in one random group in a population of 10 total initial groups. We then allow the system to evolve with no additional mutations. To have stable cheaters, we set the secretion rate of cheaters to $s_1 = 50$, whereas cooperators have $s_1 = 100$. The initial cheater takes over its group due to larger invasion fitness and remains stable. When forced through the mixer funnel (Fig. \ref{fig:geometrycontrol}b), the cheating group collides and mixes into all the cooperating groups and take over.

We explore various funnel/mixer geometries in Fig. \ref{fig:geometrycontrol}d. We describe the flow domain geometry as follows: we again fix the total width of the domain, now to $H = 25$. After a length of 16.5, we begin to narrow the width. We narrow the total width from $H$ to a minimum pinch width value $P$ over a transition length of 10. We maintain the pinch width for a length of 30 before transitioning back to the maximum width. We found varying the pinch length did not have a strong effect on results. However, we did find the effectiveness of the mixer improved with repeating the procedure. We therefore added two more identical mixer portions (Fig. \ref{fig:geometrycontrol}b) separated by a length of 3. The right length after the mixer was set to 131.5 and the simulation was run for a total time of $T = 100$.

We varied the width of the center channel $P$ compared to the maximum total width of the domain $H$. For very narrow widths, groups die out as they are forced through a narrow region. They are not able to maintain a large enough local density to maintain enough public goods and are over-polluted by waste accumulation in the narrow channel. For wider channels, the population is stable and groups are able to collide with each other. The percentage of cheaters present at the end of the run is maximized in this intermediate range of channel widths. When the channel width is larger than roughly two group widths, the groups no longer mix and cheaters do no better than they naturally would without the mixing geometry.



\subsection*{Controlling social evolution through flow}

\begin{figure*}
\centering
\includegraphics[width=0.98\textwidth]{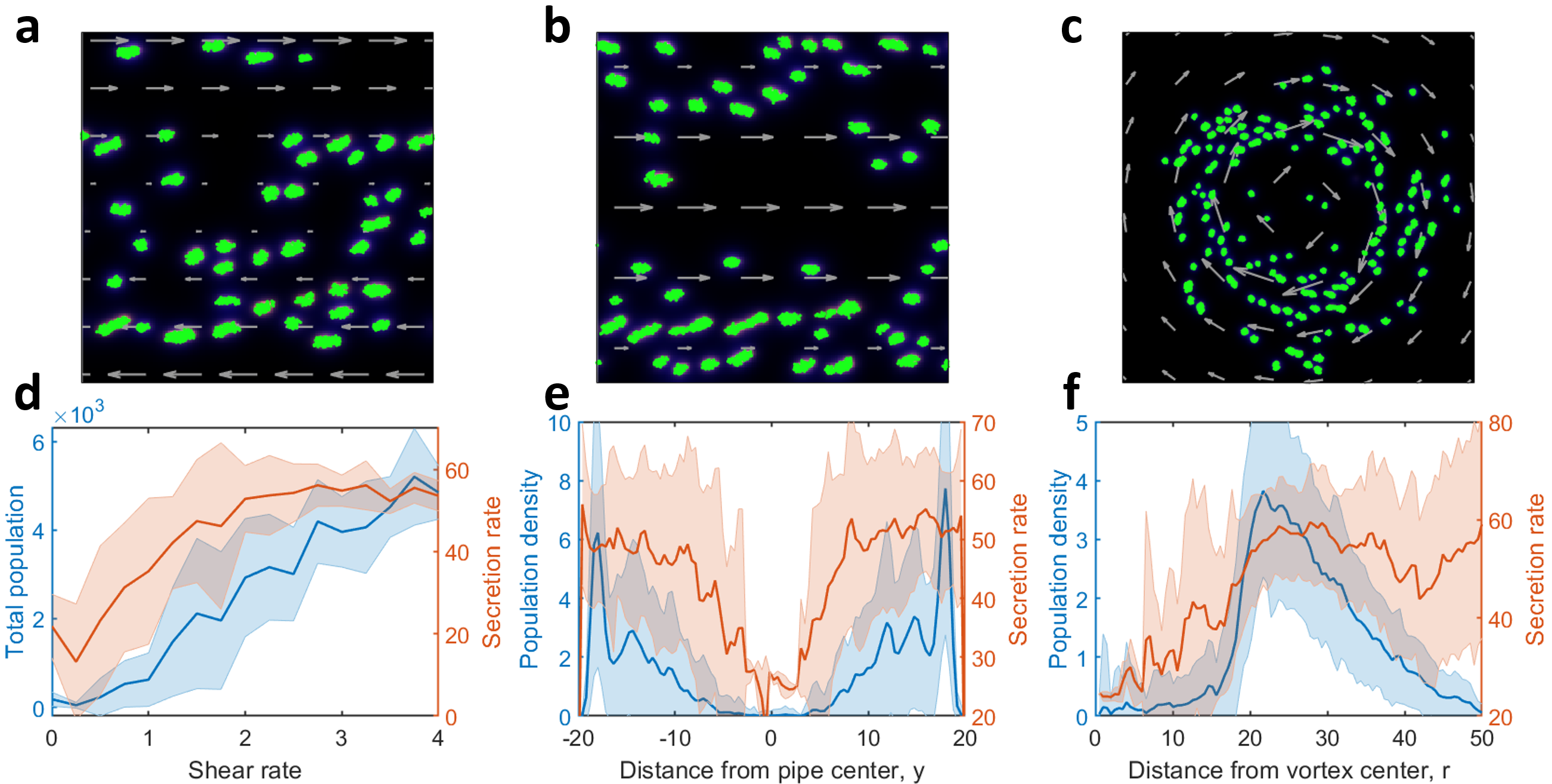}
\caption{{\bf Control via flow.} {\bf a-c,} Simulation snapshots for microbes in planar Couette ({\bf a}), Hagen-Poiseuille ({\bf b}), and Rankine vortex  ({\bf c}) flow profiles. {\bf d,} Average public good secretion rate for different shear rates in a Couette flow. A shearing flow allows groups to reproduce quicker than mutations arise. Groups with cheating mutations die out. A shearing flow therefore allows sociality to prevail, and increases the average public good secretion rate of the population. {\bf e,} Average secretion rate across a pipe with a Hagen-Poiseuille flow. Since shear rate depends on space in general, different flows can be used to control sociality locally. {\bf f,} Average secretion rate versus distance from vortex center in a Rankine vortex flow. Again we see sociality is maximized where shear is largest. Regions with high shear give rise to enhanced group fragmentation, thereby combating deleterious cheating mutations. The initial public good secretion rate was set to $s_1 = 100$, the mutation rate was set to $\sigma = 1 \times 10^{-7}$, and chemical controls were set to zero. All other parameters are as given in Table \ref{tab:parameters}. Simulations were run in a domain of size $40 \times 40$ for the Couette and Hagen-Poiseuille flow profiles and in a domain of size $100 \times 100$ for the Rankine vortex flow profile. Total time duration was $T = 50$. Shaded regions correspond to one standard deviation from the mean, represented by solid lines. Averages were taken over 10 runs for each shear rate in the Couette flow case and over 100 runs each for Hagen-Poiseuille and Rankine vortex flow. }
\label{fig:shear}
\end{figure*}

Finally, we explore controlling social behavior through fluid flow. In our previous studies \cite{uppal2018}, we found that shearing flow distorts and fragments groups quicker than the native group fragmentation rate. Therefore, by using shearing flow as a mechanism for (enhanced) group fragmentation, we can select for more cooperative populations. Populations that might otherwise fall victim to opportunistic cheating mutations can be recovered by a large shear flow as shown in Fig. \ref{fig:shear}.

We first explore a planar Couette flow to see the effects of constant shear. The flow profile for a Couette flow is given as,
\[
v(\mathbf{x}) = v_{\mathrm{max}} \frac{y}{H} \hat{x}
\]
where $v_{\mathrm{max}}$ is the maximal flow rate at boundaries of the pipe $(y=\pm H)$, $H$ is the cross-sectional radius of the pipe, flow is taken in the $\hat{x}$ direction and $y$ ranges from $-H$ to $H$. The shear rate corresponding to this flow profile is constant in space and is given by $s = \mathrm{d}v/\mathrm{d}y =  v_{\mathrm{max}}/H$. We then vary the flow rate $v_{\mathrm{max}}$ to vary the shear rate. We then see the population and average secretion rate increase with higher shear rates (Fig. \ref{fig:shear}a).

Moreover, since shear is in general spatially dependent, by controlling the flow profile, we can also control the localization of social structures. More cooperative groups tend to reside in regions of higher shear, due to larger group fragmentation rate. Laminar flow in a pipe with constant cross section can be given by the Hagen-Poiseuille law with flow profile given as,
\[
v(\mathbf{x}) = v_{\mathrm{max}} \left( 1 - \frac{y^2}{H^2} \right) \hat{x}
\]
where again $v_{\mathrm{max}}$ is the maximal flow rate, now at the center of the pipe $(y=0)$, $H$ is the cross-sectional radius of the pipe, flow is taken in the $\hat{x}$ direction and $y$ ranges from $-H$ to $H$. We then see in a pipe with a Hagen-Poiseuille flow where shear is largest at the boundaries, the average public good secretion rate is highest near the boundaries (Fig. \ref{fig:shear}b). 

We next explore social evolution in a Rankine vortex flow. The flow profile for a Rankine vortex with radius $R$ and circulation parameter $\Gamma$ is given as,
\[
v(\mathbf{x}) = \begin{cases} 
\frac{\Gamma r}{2 \pi R^2} \hat{\theta} & r \leq R \\
\frac{\Gamma}{2 \pi r} \hat{\theta} & r > R
\end{cases}
\]
where $r = x^2 + y^2$ and $\hat{\theta}$ is the angular direction. The circulation parameter $\Gamma$ corresponds to the line integral of the flow field along a closed path and has units of velocity times length. We use it here to vary the flow and shear rate of the vortex. In a Rankine vortex flow, the shear is largest in magnitude at the vortex radius $R$ after which it falls as $1/r^2$. In this case we see an annulus where sociality and population are at a maximum (Fig. \ref{fig:shear}c). 

We therefore find different flow profiles can be used to engineer systems to localize microbial cooperation to specific regions in space. 


\onecolumngrid
\begin{tcolorbox}[enhanced, width=\textwidth]
\label{box:summary}
{\bf Control through constant chemicals:} By adding a public good, microbes form denser structures and transition to no longer forming groups (Fig \ref{fig:constantcontrol}a). This enables cheaters to be stable and spread throughout the population, leading to a collapse of sociality. Adding a toxin forces microbes to secrete more public goods to compensate for the loss in fitness (Fig. \ref{fig:constantcontrol}c). However, it also lowers the population, making it more susceptible to extinction (Fig. \ref{fig:constantcontrol}b).
\\
{\bf Control through pulsed chemicals:} By alternating pulses of the public good and the toxin, we find we can increase the sociality of the population, as well as maintain a large population (Fig. \ref{fig:pulsecontrol}). Pulses of the toxin kill off weaker groups with cheaters present and select for groups that secrete more public good, harming them in the process. Subsequent pulses of public good then rescue the remaining groups and increases the population of the remaining cooperators. The next toxin pulse is then added before cheaters take over.
\\
{\bf Control with filters:} By adding channel walls, we can force groups to fragment into smaller groups. Channels with cheaters present then die off and cooperating subgroups continue to flow in their isolated channels and exit out the end of the channel without cheaters present. The channels therefore allow us to ``filter out'' cheaters from a group (Fig \ref{fig:geometrycontrol}a).
\\
{\bf Control with funnel mixers:} With a funnel geometry, we can force groups with cheaters to mix with groups of cooperators. The cheaters that mix with cooperating groups then have a larger invasion fitness and take over the cooperating groups. Therefore, a mixer can be used to reduce the overall sociality of the population (Fig \ref{fig:geometrycontrol}b). \\
{\bf Control through flow:} Flow shear can fragment social groups, thereby limiting the spread of deleterious cheating mutations. Since flow shear is spatially dependent, this can allow for localized control, with more sociality in high shear regions of the flow domain (Fig \ref{fig:shear}).  
\end{tcolorbox}
\twocolumngrid

\section*{Discussion}
We used a realistic advection-diffusion-reaction model for a microbial population to analyze the effect of external perturbations to propose possible means to enhance or inhibit the social behavior of microbial populations. Specifically, we explored introducing positive and negative chemical factors in a spatially homogeneous manner both constant and periodic in time. We also explored varying the spatial geometry and the fluid flow profile to control the fragmentation rate of microbial groups and the sociality and population of the microbes both locally and globally in space. 

We first explored adding a public good or toxin to control social behavior, and found that by adding a homogeneous field of chemicals constant in time, we can transition the microbial population from forming group patterns (spots) to forming longer stripes or simply to growing homogeneously in space (Fig. \ref{fig:constantcontrol}a). In the stripe or homogeneous states, cheaters are able to spread to the rest of the population, thus transitioning the population to a non-social state. If the added public good is sufficiently large, cheaters are stable without the presence of cooperators and can grow large in population (Fig. \ref{fig:constantcontrol}b,c). We also found that we can control the size of the population through chemicals applied constantly in time. By adding the toxin, we can force the groups to maintain higher secretion rate, but the microbial abundance is lowered and thus groups are more susceptible to extinction due to random fluctuations. 

We next investigated adding chemicals in a temporally periodic method with pulses. By alternating pulses of public good and toxin, we were able to vary the sociality and population of the system for various pulse durations (Fig. \ref{fig:pulsecontrol}b,c). For shorter pulse periods composed of roughly equal durations of public good and toxin, we were able to maintain a large secretion rate and abundance (Fig. \ref{fig:pulsecontrol}d). For pulses comprising of a larger public good duration, we were able to drive the population into an anti-social state (Fig. \ref{fig:pulsecontrol}e) and even drive the system to extinction this way (Fig. \ref{fig:pulsecontrol}f).

To control the social evolution with flow domain geometry, we investigated various filtering and mixing micro channels. By forcing groups through multiple channels, we fragmented them into smaller groups. If the original group contained some cheaters, some of the channels ended up containing these cheaters and some did not. The population in the channels with cheaters died out, whereas those in the channels without cheaters survived. 

By adding walls to construct a funnel geometry, we forced groups to mix with each other. Cheaters were then able to spread to other groups more easily, thus decreasing the sociality of the population (Fig. \ref{fig:geometrycontrol}).

Finally, we were able to use fluid flow to fragment groups and help cooperators resist cheating mutations (Fig. \ref{fig:shear}). The mechanism at play here is that shear distorts and fragments groups of microbes. This mechanism was explored in great length in our previous studies \cite{uppal2018}. Furthermore, since flow shear is, in general, spatially varying, this can be used to localize or pattern social behavior to certain select regions in space (Fig. \ref{fig:shear}e,f).

Through our first-principles physical model, we saw that the physical properties of the environment can strongly influence the social evolution of a population. As such, the sociality of the population can be seen as a mechanical phenomena dictated by the physical properties of the medium. With this viewpoint, the physical habitat can be engineered to support more or less social species, as called by medical, agro-industrial, or environmental needs.

Microbes are highly social organisms and their degree of social behavior its within-population variation has crucial evolutionary and ecological consequences. Understanding these social behaviors and the evolutionary principles underlying their origin and stability will allow us to sculpt these social structures, enhancing or inhibiting them spatially and temporally, according to our practical needs.

Although we tried to retain the most relevant features in our model, we made some simplifying assumptions which in certain circumstances may lead to incorrect conclusions. First, since microorganisms typically live in a low Reynolds number environment \cite{purcell1977life}, we have chosen to neglect the inertia of microorganisms. In reality the microbes also influence the flow around them. This effect will be particularly significant for a dense population. We have also neglected the size of the microbes and treat them as point particles. This effect also becomes significant in the dense population limit or when microbes aggregate and stick to each other or to surfaces via extracellular polymeric substances. These assumptions may also affect our results in controlling with geometry. For example, finite size and sticking effects may lead to clogging or jamming in our microfluidic channels. It has been shown that porous environments can influence competition between biofilms where large growth rates may lead to the redirection of nutrients received by competing colonies \cite{coyte2017microbial}.

Our model is fairly general and could perhaps be extended to study the social evolution of other systems such as models of cancer \cite{archetti2019cooperation,acar2020exploiting,dawson2018social}. A number of additional assumptions would need to be made or modified in order to do this. For example, we consider here Monod growth kinetics for microbes. Tumor growth is more typically modeled with logistic, Gompertz, or Von Bertalanffy models \cite{benzekry2014classical}. However, these variations still mainly have an exponential growth phase followed by a saturating growth plataeu and are expected to give similar results \cite{koziol2020different}. We have also ignored the active response of microorganisms to their surrounding chemical gradients and model them as simple Brownian particles. Chemotaxis and collective migration would also be relevant to add to models describing cancer growth and metastasis \cite{roussos2011chemotaxis,cheung2016collective,cheung2016polyclonal}. Additional features such as direct physical cell-cell interactions may become more relevant for models of cancer and models such as Cellular Potts \cite{graner1992simulation,szabo2013cellular} or active vertex type models \cite{barton2017active,koride2018epithelial} would be more suitable to accurately represent these interactions. 


We hope that our findings will inspire further research into controlling social evolution in populations. Interesting future studies may also investigate combining the various mechanisms offered in this study. Feedback control mechanisms can also be investigated to more optimally regulate the sociality and population of the system. It would also be interesting to explore chemicals added in a spatially dependent manner. This has the potential to pinpoint the localization of social interactions within a broader system. However, as additional chemicals disperse, it could also give rise to a gradient of social states across space. Here we studied the social evolution of a single trait. It would also be interesting to explore the influence of our control strategies on multi-trait systems and the evolution of division of labor \cite{uppal2020evolution}. 

\section*{Methods}
Our agent based stochastic simulations are guided by equations (\ref{eq:microbes}-\ref{eq:waste}). In our model, microbes can secrete a cooperative public good and a waste compound. These molecules diffuse, advect, and decay according to equations \ref{eq:goods} and \ref{eq:waste} and are updated via a finite element scheme as detailed in Appendix \ref{app:model}. We simulate the evolution of bacteria as discrete agents in a two dimensional flowing channel, using periodic boundary conditions for the left and right walls and Neumann boundary conditions on the top and bottom walls, unless stated otherwise. 

Our simulation algorithm is as follows: at each time interval, $\Delta t$, the microbes (1) diffuse via a random walk, with step size $\delta = \sqrt{4 d_b \Delta t}$ derived from the diffusion constant and a bias dependent on the flow velocity, $ \vec{v} \Delta t$, (2) secrete chemicals locally that then diffuse and advect using a finite element scheme, and (3) reproduce or die with a probability dependent on their local fitness given by $f = \Delta t \left[ \alpha_1 \frac{c_1}{c_1 + k_1} - \alpha_2 \frac{c_2}{c_2 + k_2} - \beta_1 s_1 \right]$. If $f$ is negative, the microbes die with probability 1, if $f$ is between 0 and 1 they reproduce with probability $f$. Fitness constants and time step are always sufficiently small so that $|f| < 1$. Upon reproduction, random mutations may alter the secretion rate of the public good --and thus the reproduction rate-- of the microbes. Mutations occur with probability $\mu$ and can change the secretion rate by a random number between $0$ and $s_1$. The secretion rate is assumed to be heritable, and constant in time. We should not expect the discrete simulations to perfectly be described by the continuous set of partial differential equations. Nevertheless, the continuous system of equations do allow us to obtain relevant quantities such as group size and group fragmentation rate.  Our simulation code is written in C++ using the open source deal.II library for implementing the finite element fields \cite{bangerth2007deal} and is available on Github: \url{https://github.com/garyuppal/MicrobeSimulator.git}. Videos of simulations are also provided as supplementary files. 

\begin{table}
\centering
\begin{tabular}{llll}
Parameter & Definition & Values & Sources \\
$d_b$ 	 		& Bacteria diffusion constant 		& 0.4 & \cite{Kim1996} \\
$d_1$	 	 	& Public good diffusion constant 	& 5 & \cite{Ma2005}\\
$d_2$	 		& Waste diffusion constant 			& 15 & \cite{Ma2005}\\
$|v|$	 		& Flow rate 			& 0 - 80 & \cite{rusconi2015microbes,busscher2006microbial}\\
$\lambda_1$		& Public good decay constant 		& 50\\
$\lambda_2$		& Waste decay constant 				& 15\\
$k_1$ 			& Public good saturation 			& 0.01\\
$k_2$ 		    & Waste saturation 					& 0.1\\
$s_1$ 		 	& Public good secretion rate		& 0 - 100 \\
$s_2$ 		 	& Waste secretion rate 				& 100 \\
$\alpha_1$ 			& Benefit of public good 			& 75 & \cite{gibson2018distribution} \\
$\alpha_2$ 			& Harm of waste compound 			& 80 & \cite{gibson2018distribution} \\
$\beta_1$ 			& Cost of secretion 				& 0.02 & \cite{gibson2018distribution} \\
$\sigma$ 			& Mutation rate 					& $10^{-7}$ - $10^{-6}$ & \cite{drake1998rates} \\
\end{tabular}
\caption{Summary of general system parameters and the values used in our simulations. Values for parameters were chosen to fit suitable ranges given from sources. Further details on parameter choices are given in Appendix \ref{app:model}. } \label{tab:parameters}
\end{table}

A summary of the system parameters is given in Table \ref{tab:parameters}, along with typical ranges for their values used in the simulations and sources from which these values were obtained. The ranges of parameters were kept to match those observed empirically for diffusion rates \cite{Kim1996,Ma2005}, flow rates \cite{rusconi2015microbes}, and mutation rates \cite{drake1998rates}. Secretion rates can be rescaled along with saturation and decay constants. The relevant quantity is then the range of growth rates. These were kept consistent with empirically observed conditions for growth rates \cite{gibson2018distribution}. We also restrict $\alpha_1 < \alpha_2$ since otherwise, when adding a public good externally, cheaters pay no cost $(\beta_1 s_1 = 0)$ and the fitness could become positive for a dense population, leading to an infinite population. Other constraints on existence and stability are derived in our Turing analysis (see Appendix \ref{app:turing}). Additional details of our model implementation and choice of system parameters are provided in Appendix \ref{app:model}.

\section*{Supplementary materials}
\begin{enumerate}
    \item {\bf Supplementary video.} Simulation video with constant public goods added, leading to take over by cheaters. The initial public good secretion rate was set to $s_1 = 100$, the mutation rate was set to $\sigma = 1 \times 10^{-7}$, and flow rate was set to zero. The public good strength is $\mu_1 = 0.3$ and the toxin strength is $\mu_2 = 0$. All other parameters are as given in Table \ref{tab:parameters}. The simulation domain is of size $40 \times 40$ and has total duration $T = 50$.   \label{itm:const_pg}
    \item {\bf Supplementary video.} Simulation video with constant toxins added, causing emerging cheaters to die out and allowing for stable cooperation. The initial public good secretion rate was set to $s_1 = 100$, the mutation rate was set to $\sigma = 1 \times 10^{-7}$, and flow rate was set to zero. The public good strength is $\mu_1 = 0$ and the toxin strength is $\mu_2 = 0.4$. All other parameters are as given in Table \ref{tab:parameters}. The simulation domain is of size $40 \times 40$ and has total duration $T = 50$.  \label{itm:const_toxin}
    \item {\bf Supplementary video.} Simulation video with chemical pulses leading to enhanced group fragmentation and stable cooperation. The initial public good secretion rate was set to $s_1 = 100$, the mutation rate was set to $\sigma = 1 \times 10^{-7}$, and flow rate was set to zero. The public good amplitude is $A_1 = 0.5$ and the toxin amplitude is $A_2 = 0.3$. Pulse durations are $\tau_1 = 0.7$ for the public good and $\tau_2 = 0.7$ for the toxin, giving $T = \tau_1 + \tau_2 = 1.4$ for the full period. All other parameters are as given in Table \ref{tab:parameters}. The simulation domain is of size $40 \times 40$ and has total duration $T = 50$.  \label{itm:pulses}
\end{enumerate}


\section*{Author Contributions}
GU and DCV formulated the problem. GU implemented the software and carried out numerical experiments. GU and DCV designed and carried out the theory. GU and DCV wrote the paper.

\section*{Acknowledgments}

This study was partially supported by NSF under award number CBET-1805157.

\section*{Declaration of interest}
The authors declare no competing interests.

\bibliography{bibliography}

\begin{thebibliography}{75}%
\makeatletter
\providecommand \@ifxundefined [1]{%
 \@ifx{#1\undefined}
}%
\providecommand \@ifnum [1]{%
 \ifnum #1\expandafter \@firstoftwo
 \else \expandafter \@secondoftwo
 \fi
}%
\providecommand \@ifx [1]{%
 \ifx #1\expandafter \@firstoftwo
 \else \expandafter \@secondoftwo
 \fi
}%
\providecommand \natexlab [1]{#1}%
\providecommand \enquote  [1]{``#1''}%
\providecommand \bibnamefont  [1]{#1}%
\providecommand \bibfnamefont [1]{#1}%
\providecommand \citenamefont [1]{#1}%
\providecommand \href@noop [0]{\@secondoftwo}%
\providecommand \href [0]{\begingroup \@sanitize@url \@href}%
\providecommand \@href[1]{\@@startlink{#1}\@@href}%
\providecommand \@@href[1]{\endgroup#1\@@endlink}%
\providecommand \@sanitize@url [0]{\catcode `\\12\catcode `\$12\catcode `\&12\catcode `\#12\catcode `\^12\catcode `\_12\catcode `\%12\relax}%
\providecommand \@@startlink[1]{}%
\providecommand \@@endlink[0]{}%
\providecommand \url  [0]{\begingroup\@sanitize@url \@url }%
\providecommand \@url [1]{\endgroup\@href {#1}{\urlprefix }}%
\providecommand \urlprefix  [0]{URL }%
\providecommand \Eprint [0]{\href }%
\providecommand \doibase [0]{http://dx.doi.org/}%
\providecommand \selectlanguage [0]{\@gobble}%
\providecommand \bibinfo  [0]{\@secondoftwo}%
\providecommand \bibfield  [0]{\@secondoftwo}%
\providecommand \translation [1]{[#1]}%
\providecommand \BibitemOpen [0]{}%
\providecommand \bibitemStop [0]{}%
\providecommand \bibitemNoStop [0]{.\EOS\space}%
\providecommand \EOS [0]{\spacefactor3000\relax}%
\providecommand \BibitemShut  [1]{\csname bibitem#1\endcsname}%
\let\auto@bib@innerbib\@empty
\bibitem [{\citenamefont {Pan}\ \emph {et~al.}(2012)\citenamefont {Pan}, \citenamefont {Dam},\ and\ \citenamefont {Sen}}]{pan2012composting}%
  \BibitemOpen
  \bibfield  {author} {\bibinfo {author} {\bibfnamefont {I.}~\bibnamefont {Pan}}, \bibinfo {author} {\bibfnamefont {B.}~\bibnamefont {Dam}}, \ and\ \bibinfo {author} {\bibfnamefont {S.}~\bibnamefont {Sen}},\ }\href@noop {} {\bibfield  {journal} {\bibinfo  {journal} {3 Biotech}\ }\textbf {\bibinfo {volume} {2}},\ \bibinfo {pages} {127} (\bibinfo {year} {2012})}\BibitemShut {NoStop}%
\bibitem [{\citenamefont {Cohen}(2001)}]{cohen2001biofiltration}%
  \BibitemOpen
  \bibfield  {author} {\bibinfo {author} {\bibfnamefont {Y.}~\bibnamefont {Cohen}},\ }\href@noop {} {\bibfield  {journal} {\bibinfo  {journal} {Bioresource technology}\ }\textbf {\bibinfo {volume} {77}},\ \bibinfo {pages} {257} (\bibinfo {year} {2001})}\BibitemShut {NoStop}%
\bibitem [{\citenamefont {Zhang}\ \emph {et~al.}(2015)\citenamefont {Zhang}, \citenamefont {Li}, \citenamefont {Gan}, \citenamefont {Zhou}, \citenamefont {Xu},\ and\ \citenamefont {Li}}]{zhang2015impacts}%
  \BibitemOpen
  \bibfield  {author} {\bibinfo {author} {\bibfnamefont {Y.-J.}\ \bibnamefont {Zhang}}, \bibinfo {author} {\bibfnamefont {S.}~\bibnamefont {Li}}, \bibinfo {author} {\bibfnamefont {R.-Y.}\ \bibnamefont {Gan}}, \bibinfo {author} {\bibfnamefont {T.}~\bibnamefont {Zhou}}, \bibinfo {author} {\bibfnamefont {D.-P.}\ \bibnamefont {Xu}}, \ and\ \bibinfo {author} {\bibfnamefont {H.-B.}\ \bibnamefont {Li}},\ }\href@noop {} {\bibfield  {journal} {\bibinfo  {journal} {International journal of molecular sciences}\ }\textbf {\bibinfo {volume} {16}},\ \bibinfo {pages} {7493} (\bibinfo {year} {2015})}\BibitemShut {NoStop}%
\bibitem [{\citenamefont {Cogdell}\ \emph {et~al.}(1999)\citenamefont {Cogdell}, \citenamefont {Isaacs}, \citenamefont {Howard}, \citenamefont {McLuskey}, \citenamefont {Fraser},\ and\ \citenamefont {Prince}}]{cogdell1999photosynthetic}%
  \BibitemOpen
  \bibfield  {author} {\bibinfo {author} {\bibfnamefont {R.~J.}\ \bibnamefont {Cogdell}}, \bibinfo {author} {\bibfnamefont {N.~W.}\ \bibnamefont {Isaacs}}, \bibinfo {author} {\bibfnamefont {T.~D.}\ \bibnamefont {Howard}}, \bibinfo {author} {\bibfnamefont {K.}~\bibnamefont {McLuskey}}, \bibinfo {author} {\bibfnamefont {N.~J.}\ \bibnamefont {Fraser}}, \ and\ \bibinfo {author} {\bibfnamefont {S.~M.}\ \bibnamefont {Prince}},\ }\href@noop {} {\bibfield  {journal} {\bibinfo  {journal} {Journal of bacteriology}\ }\textbf {\bibinfo {volume} {181}},\ \bibinfo {pages} {3869} (\bibinfo {year} {1999})}\BibitemShut {NoStop}%
\bibitem [{\citenamefont {Xia}\ \emph {et~al.}(2018)\citenamefont {Xia}, \citenamefont {Wen}, \citenamefont {Zhang},\ and\ \citenamefont {Yang}}]{xia2018diversity}%
  \BibitemOpen
  \bibfield  {author} {\bibinfo {author} {\bibfnamefont {Y.}~\bibnamefont {Xia}}, \bibinfo {author} {\bibfnamefont {X.}~\bibnamefont {Wen}}, \bibinfo {author} {\bibfnamefont {B.}~\bibnamefont {Zhang}}, \ and\ \bibinfo {author} {\bibfnamefont {Y.}~\bibnamefont {Yang}},\ }\href@noop {} {\bibfield  {journal} {\bibinfo  {journal} {Biotechnology advances}\ }\textbf {\bibinfo {volume} {36}},\ \bibinfo {pages} {1038} (\bibinfo {year} {2018})}\BibitemShut {NoStop}%
\bibitem [{\citenamefont {Dor{\'e}}\ and\ \citenamefont {Blotti{\`e}re}(2015)}]{dore2015influence}%
  \BibitemOpen
  \bibfield  {author} {\bibinfo {author} {\bibfnamefont {J.}~\bibnamefont {Dor{\'e}}}\ and\ \bibinfo {author} {\bibfnamefont {H.}~\bibnamefont {Blotti{\`e}re}},\ }\href@noop {} {\bibfield  {journal} {\bibinfo  {journal} {Current opinion in biotechnology}\ }\textbf {\bibinfo {volume} {32}},\ \bibinfo {pages} {195} (\bibinfo {year} {2015})}\BibitemShut {NoStop}%
\bibitem [{\citenamefont {Lai}\ \emph {et~al.}(2009)\citenamefont {Lai}, \citenamefont {Tremblay},\ and\ \citenamefont {D{\'e}ziel}}]{lai2009swarming}%
  \BibitemOpen
  \bibfield  {author} {\bibinfo {author} {\bibfnamefont {S.}~\bibnamefont {Lai}}, \bibinfo {author} {\bibfnamefont {J.}~\bibnamefont {Tremblay}}, \ and\ \bibinfo {author} {\bibfnamefont {E.}~\bibnamefont {D{\'e}ziel}},\ }\href@noop {} {\bibfield  {journal} {\bibinfo  {journal} {Environmental microbiology}\ }\textbf {\bibinfo {volume} {11}},\ \bibinfo {pages} {126} (\bibinfo {year} {2009})}\BibitemShut {NoStop}%
\bibitem [{\citenamefont {Stewart}(2002)}]{stewart2002mechanisms}%
  \BibitemOpen
  \bibfield  {author} {\bibinfo {author} {\bibfnamefont {P.~S.}\ \bibnamefont {Stewart}},\ }\href@noop {} {\bibfield  {journal} {\bibinfo  {journal} {International journal of medical microbiology}\ }\textbf {\bibinfo {volume} {292}},\ \bibinfo {pages} {107} (\bibinfo {year} {2002})}\BibitemShut {NoStop}%
\bibitem [{\citenamefont {Cox}\ and\ \citenamefont {Wright}(2013)}]{cox2013intrinsic}%
  \BibitemOpen
  \bibfield  {author} {\bibinfo {author} {\bibfnamefont {G.}~\bibnamefont {Cox}}\ and\ \bibinfo {author} {\bibfnamefont {G.~D.}\ \bibnamefont {Wright}},\ }\href@noop {} {\bibfield  {journal} {\bibinfo  {journal} {International Journal of Medical Microbiology}\ }\textbf {\bibinfo {volume} {303}},\ \bibinfo {pages} {287} (\bibinfo {year} {2013})}\BibitemShut {NoStop}%
\bibitem [{\citenamefont {de~Carvalho}(2018)}]{de2018marine}%
  \BibitemOpen
  \bibfield  {author} {\bibinfo {author} {\bibfnamefont {C.~C.}\ \bibnamefont {de~Carvalho}},\ }\href@noop {} {\bibfield  {journal} {\bibinfo  {journal} {Frontiers in Marine Science}\ }\textbf {\bibinfo {volume} {5}},\ \bibinfo {pages} {126} (\bibinfo {year} {2018})}\BibitemShut {NoStop}%
\bibitem [{\citenamefont {Schultz}\ \emph {et~al.}(2011)\citenamefont {Schultz}, \citenamefont {Bendick}, \citenamefont {Holm},\ and\ \citenamefont {Hertel}}]{schultz2011economic}%
  \BibitemOpen
  \bibfield  {author} {\bibinfo {author} {\bibfnamefont {M.}~\bibnamefont {Schultz}}, \bibinfo {author} {\bibfnamefont {J.}~\bibnamefont {Bendick}}, \bibinfo {author} {\bibfnamefont {E.}~\bibnamefont {Holm}}, \ and\ \bibinfo {author} {\bibfnamefont {W.}~\bibnamefont {Hertel}},\ }\href@noop {} {\bibfield  {journal} {\bibinfo  {journal} {Biofouling}\ }\textbf {\bibinfo {volume} {27}},\ \bibinfo {pages} {87} (\bibinfo {year} {2011})}\BibitemShut {NoStop}%
\bibitem [{\citenamefont {Joyce}\ and\ \citenamefont {Pollard}(2009)}]{joyce2009microenvironmental}%
  \BibitemOpen
  \bibfield  {author} {\bibinfo {author} {\bibfnamefont {J.~A.}\ \bibnamefont {Joyce}}\ and\ \bibinfo {author} {\bibfnamefont {J.~W.}\ \bibnamefont {Pollard}},\ }\href@noop {} {\bibfield  {journal} {\bibinfo  {journal} {Nature reviews cancer}\ }\textbf {\bibinfo {volume} {9}},\ \bibinfo {pages} {239} (\bibinfo {year} {2009})}\BibitemShut {NoStop}%
\bibitem [{\citenamefont {Palkov{\'a}}(2004)}]{palkova2004multicellular}%
  \BibitemOpen
  \bibfield  {author} {\bibinfo {author} {\bibfnamefont {Z.}~\bibnamefont {Palkov{\'a}}},\ }\href@noop {} {\bibfield  {journal} {\bibinfo  {journal} {EMBO reports}\ }\textbf {\bibinfo {volume} {5}},\ \bibinfo {pages} {470} (\bibinfo {year} {2004})}\BibitemShut {NoStop}%
\bibitem [{\citenamefont {Kim}\ \emph {et~al.}(2016)\citenamefont {Kim}, \citenamefont {Ingremeau}, \citenamefont {Zhao}, \citenamefont {Bassler},\ and\ \citenamefont {Stone}}]{kim2016local}%
  \BibitemOpen
  \bibfield  {author} {\bibinfo {author} {\bibfnamefont {M.~K.}\ \bibnamefont {Kim}}, \bibinfo {author} {\bibfnamefont {F.}~\bibnamefont {Ingremeau}}, \bibinfo {author} {\bibfnamefont {A.}~\bibnamefont {Zhao}}, \bibinfo {author} {\bibfnamefont {B.~L.}\ \bibnamefont {Bassler}}, \ and\ \bibinfo {author} {\bibfnamefont {H.~A.}\ \bibnamefont {Stone}},\ }\href@noop {} {\bibfield  {journal} {\bibinfo  {journal} {Nature microbiology}\ }\textbf {\bibinfo {volume} {1}},\ \bibinfo {pages} {1} (\bibinfo {year} {2016})}\BibitemShut {NoStop}%
\bibitem [{\citenamefont {Stoodley}(2016)}]{stoodley2016biofilms}%
  \BibitemOpen
  \bibfield  {author} {\bibinfo {author} {\bibfnamefont {P.}~\bibnamefont {Stoodley}},\ }\href@noop {} {\bibfield  {journal} {\bibinfo  {journal} {Nature microbiology}\ }\textbf {\bibinfo {volume} {1}},\ \bibinfo {pages} {15012} (\bibinfo {year} {2016})}\BibitemShut {NoStop}%
\bibitem [{\citenamefont {Wechsler}\ \emph {et~al.}(2019)\citenamefont {Wechsler}, \citenamefont {K{\"u}mmerli},\ and\ \citenamefont {Dobay}}]{wechsler2019understanding}%
  \BibitemOpen
  \bibfield  {author} {\bibinfo {author} {\bibfnamefont {T.}~\bibnamefont {Wechsler}}, \bibinfo {author} {\bibfnamefont {R.}~\bibnamefont {K{\"u}mmerli}}, \ and\ \bibinfo {author} {\bibfnamefont {A.}~\bibnamefont {Dobay}},\ }\href@noop {} {\bibfield  {journal} {\bibinfo  {journal} {Journal of evolutionary biology}\ }\textbf {\bibinfo {volume} {32}},\ \bibinfo {pages} {412} (\bibinfo {year} {2019})}\BibitemShut {NoStop}%
\bibitem [{\citenamefont {Uppal}\ and\ \citenamefont {Vural}(2018)}]{uppal2018}%
  \BibitemOpen
  \bibfield  {author} {\bibinfo {author} {\bibfnamefont {G.}~\bibnamefont {Uppal}}\ and\ \bibinfo {author} {\bibfnamefont {D.~C.}\ \bibnamefont {Vural}},\ }\href@noop {} {\bibfield  {journal} {\bibinfo  {journal} {eLife}\ }\textbf {\bibinfo {volume} {7}},\ \bibinfo {pages} {e34862} (\bibinfo {year} {2018})}\BibitemShut {NoStop}%
\bibitem [{\citenamefont {Uppal}\ and\ \citenamefont {Vural}(2020)}]{uppal2020evolution}%
  \BibitemOpen
  \bibfield  {author} {\bibinfo {author} {\bibfnamefont {G.}~\bibnamefont {Uppal}}\ and\ \bibinfo {author} {\bibfnamefont {D.~C.}\ \bibnamefont {Vural}},\ }\href@noop {} {\bibfield  {journal} {\bibinfo  {journal} {Journal of Evolutionary Biology}\ }\textbf {\bibinfo {volume} {33}},\ \bibinfo {pages} {256} (\bibinfo {year} {2020})}\BibitemShut {NoStop}%
\bibitem [{\citenamefont {Brockhurst}\ \emph {et~al.}(2007)\citenamefont {Brockhurst}, \citenamefont {Buckling},\ and\ \citenamefont {Gardner}}]{brockhurst2007cooperation}%
  \BibitemOpen
  \bibfield  {author} {\bibinfo {author} {\bibfnamefont {M.~A.}\ \bibnamefont {Brockhurst}}, \bibinfo {author} {\bibfnamefont {A.}~\bibnamefont {Buckling}}, \ and\ \bibinfo {author} {\bibfnamefont {A.}~\bibnamefont {Gardner}},\ }\href@noop {} {\bibfield  {journal} {\bibinfo  {journal} {Current Biology}\ }\textbf {\bibinfo {volume} {17}},\ \bibinfo {pages} {761} (\bibinfo {year} {2007})}\BibitemShut {NoStop}%
\bibitem [{\citenamefont {Archetti}\ and\ \citenamefont {Pienta}(2019)}]{archetti2019cooperation}%
  \BibitemOpen
  \bibfield  {author} {\bibinfo {author} {\bibfnamefont {M.}~\bibnamefont {Archetti}}\ and\ \bibinfo {author} {\bibfnamefont {K.~J.}\ \bibnamefont {Pienta}},\ }\href@noop {} {\bibfield  {journal} {\bibinfo  {journal} {Nature Reviews Cancer}\ }\textbf {\bibinfo {volume} {19}},\ \bibinfo {pages} {110} (\bibinfo {year} {2019})}\BibitemShut {NoStop}%
\bibitem [{\citenamefont {Cavaliere}\ \emph {et~al.}(2017)\citenamefont {Cavaliere}, \citenamefont {Feng}, \citenamefont {Soyer},\ and\ \citenamefont {Jim{\'e}nez}}]{cavaliere2017cooperation}%
  \BibitemOpen
  \bibfield  {author} {\bibinfo {author} {\bibfnamefont {M.}~\bibnamefont {Cavaliere}}, \bibinfo {author} {\bibfnamefont {S.}~\bibnamefont {Feng}}, \bibinfo {author} {\bibfnamefont {O.~S.}\ \bibnamefont {Soyer}}, \ and\ \bibinfo {author} {\bibfnamefont {J.~I.}\ \bibnamefont {Jim{\'e}nez}},\ }\href@noop {} {\bibfield  {journal} {\bibinfo  {journal} {Environmental microbiology}\ }\textbf {\bibinfo {volume} {19}},\ \bibinfo {pages} {2949} (\bibinfo {year} {2017})}\BibitemShut {NoStop}%
\bibitem [{\citenamefont {Sheth}\ \emph {et~al.}(2016)\citenamefont {Sheth}, \citenamefont {Cabral}, \citenamefont {Chen},\ and\ \citenamefont {Wang}}]{sheth2016manipulating}%
  \BibitemOpen
  \bibfield  {author} {\bibinfo {author} {\bibfnamefont {R.~U.}\ \bibnamefont {Sheth}}, \bibinfo {author} {\bibfnamefont {V.}~\bibnamefont {Cabral}}, \bibinfo {author} {\bibfnamefont {S.~P.}\ \bibnamefont {Chen}}, \ and\ \bibinfo {author} {\bibfnamefont {H.~H.}\ \bibnamefont {Wang}},\ }\href@noop {} {\bibfield  {journal} {\bibinfo  {journal} {Trends in Genetics}\ }\textbf {\bibinfo {volume} {32}},\ \bibinfo {pages} {189} (\bibinfo {year} {2016})}\BibitemShut {NoStop}%
\bibitem [{\citenamefont {Roberfroid}\ \emph {et~al.}(2010)\citenamefont {Roberfroid}, \citenamefont {Gibson}, \citenamefont {Hoyles}, \citenamefont {McCartney}, \citenamefont {Rastall}, \citenamefont {Rowland}, \citenamefont {Wolvers}, \citenamefont {Watzl}, \citenamefont {Szajewska}, \citenamefont {Stahl} \emph {et~al.}}]{roberfroid2010prebiotic}%
  \BibitemOpen
  \bibfield  {author} {\bibinfo {author} {\bibfnamefont {M.}~\bibnamefont {Roberfroid}}, \bibinfo {author} {\bibfnamefont {G.~R.}\ \bibnamefont {Gibson}}, \bibinfo {author} {\bibfnamefont {L.}~\bibnamefont {Hoyles}}, \bibinfo {author} {\bibfnamefont {A.~L.}\ \bibnamefont {McCartney}}, \bibinfo {author} {\bibfnamefont {R.}~\bibnamefont {Rastall}}, \bibinfo {author} {\bibfnamefont {I.}~\bibnamefont {Rowland}}, \bibinfo {author} {\bibfnamefont {D.}~\bibnamefont {Wolvers}}, \bibinfo {author} {\bibfnamefont {B.}~\bibnamefont {Watzl}}, \bibinfo {author} {\bibfnamefont {H.}~\bibnamefont {Szajewska}}, \bibinfo {author} {\bibfnamefont {B.}~\bibnamefont {Stahl}},  \emph {et~al.},\ }\href@noop {} {\bibfield  {journal} {\bibinfo  {journal} {British Journal of Nutrition}\ }\textbf {\bibinfo {volume} {104}},\ \bibinfo {pages} {S1} (\bibinfo {year} {2010})}\BibitemShut {NoStop}%
\bibitem [{\citenamefont {Bouhnik}\ \emph {et~al.}(2004)\citenamefont {Bouhnik}, \citenamefont {Raskine}, \citenamefont {Simoneau}, \citenamefont {Vicaut}, \citenamefont {Neut}, \citenamefont {Flouri{\'e}}, \citenamefont {Brouns},\ and\ \citenamefont {Bornet}}]{bouhnik2004capacity}%
  \BibitemOpen
  \bibfield  {author} {\bibinfo {author} {\bibfnamefont {Y.}~\bibnamefont {Bouhnik}}, \bibinfo {author} {\bibfnamefont {L.}~\bibnamefont {Raskine}}, \bibinfo {author} {\bibfnamefont {G.}~\bibnamefont {Simoneau}}, \bibinfo {author} {\bibfnamefont {E.}~\bibnamefont {Vicaut}}, \bibinfo {author} {\bibfnamefont {C.}~\bibnamefont {Neut}}, \bibinfo {author} {\bibfnamefont {B.}~\bibnamefont {Flouri{\'e}}}, \bibinfo {author} {\bibfnamefont {F.}~\bibnamefont {Brouns}}, \ and\ \bibinfo {author} {\bibfnamefont {F.~R.}\ \bibnamefont {Bornet}},\ }\href@noop {} {\bibfield  {journal} {\bibinfo  {journal} {The American journal of clinical nutrition}\ }\textbf {\bibinfo {volume} {80}},\ \bibinfo {pages} {1658} (\bibinfo {year} {2004})}\BibitemShut {NoStop}%
\bibitem [{\citenamefont {Kohanski}\ \emph {et~al.}(2010)\citenamefont {Kohanski}, \citenamefont {Dwyer},\ and\ \citenamefont {Collins}}]{kohanski2010antibiotics}%
  \BibitemOpen
  \bibfield  {author} {\bibinfo {author} {\bibfnamefont {M.~A.}\ \bibnamefont {Kohanski}}, \bibinfo {author} {\bibfnamefont {D.~J.}\ \bibnamefont {Dwyer}}, \ and\ \bibinfo {author} {\bibfnamefont {J.~J.}\ \bibnamefont {Collins}},\ }\href@noop {} {\bibfield  {journal} {\bibinfo  {journal} {Nature Reviews Microbiology}\ }\textbf {\bibinfo {volume} {8}},\ \bibinfo {pages} {423} (\bibinfo {year} {2010})}\BibitemShut {NoStop}%
\bibitem [{\citenamefont {Robinson}\ and\ \citenamefont {Young}(2010)}]{robinson2010antibiotic}%
  \BibitemOpen
  \bibfield  {author} {\bibinfo {author} {\bibfnamefont {C.~J.}\ \bibnamefont {Robinson}}\ and\ \bibinfo {author} {\bibfnamefont {V.~B.}\ \bibnamefont {Young}},\ }\href@noop {} {\bibfield  {journal} {\bibinfo  {journal} {Gut microbes}\ }\textbf {\bibinfo {volume} {1}},\ \bibinfo {pages} {279} (\bibinfo {year} {2010})}\BibitemShut {NoStop}%
\bibitem [{\citenamefont {Murray}\ \emph {et~al.}(2018)\citenamefont {Murray}, \citenamefont {Zhang}, \citenamefont {Yin}, \citenamefont {Zhang}, \citenamefont {Buckling}, \citenamefont {Snape},\ and\ \citenamefont {Gaze}}]{murray2018novel}%
  \BibitemOpen
  \bibfield  {author} {\bibinfo {author} {\bibfnamefont {A.~K.}\ \bibnamefont {Murray}}, \bibinfo {author} {\bibfnamefont {L.}~\bibnamefont {Zhang}}, \bibinfo {author} {\bibfnamefont {X.}~\bibnamefont {Yin}}, \bibinfo {author} {\bibfnamefont {T.}~\bibnamefont {Zhang}}, \bibinfo {author} {\bibfnamefont {A.}~\bibnamefont {Buckling}}, \bibinfo {author} {\bibfnamefont {J.}~\bibnamefont {Snape}}, \ and\ \bibinfo {author} {\bibfnamefont {W.~H.}\ \bibnamefont {Gaze}},\ }\href@noop {} {\bibfield  {journal} {\bibinfo  {journal} {MBio}\ }\textbf {\bibinfo {volume} {9}},\ \bibinfo {pages} {e00969} (\bibinfo {year} {2018})}\BibitemShut {NoStop}%
\bibitem [{\citenamefont {McGranahan}\ and\ \citenamefont {Swanton}(2015)}]{mcgranahan2015biological}%
  \BibitemOpen
  \bibfield  {author} {\bibinfo {author} {\bibfnamefont {N.}~\bibnamefont {McGranahan}}\ and\ \bibinfo {author} {\bibfnamefont {C.}~\bibnamefont {Swanton}},\ }\href@noop {} {\bibfield  {journal} {\bibinfo  {journal} {Cancer cell}\ }\textbf {\bibinfo {volume} {27}},\ \bibinfo {pages} {15} (\bibinfo {year} {2015})}\BibitemShut {NoStop}%
\bibitem [{\citenamefont {de~la Fuente-N{\'u}{\~n}ez}\ \emph {et~al.}(2013)\citenamefont {de~la Fuente-N{\'u}{\~n}ez}, \citenamefont {Reffuveille}, \citenamefont {Fern{\'a}ndez},\ and\ \citenamefont {Hancock}}]{de2013bacterial}%
  \BibitemOpen
  \bibfield  {author} {\bibinfo {author} {\bibfnamefont {C.}~\bibnamefont {de~la Fuente-N{\'u}{\~n}ez}}, \bibinfo {author} {\bibfnamefont {F.}~\bibnamefont {Reffuveille}}, \bibinfo {author} {\bibfnamefont {L.}~\bibnamefont {Fern{\'a}ndez}}, \ and\ \bibinfo {author} {\bibfnamefont {R.~E.}\ \bibnamefont {Hancock}},\ }\href@noop {} {\bibfield  {journal} {\bibinfo  {journal} {Current opinion in microbiology}\ }\textbf {\bibinfo {volume} {16}},\ \bibinfo {pages} {580} (\bibinfo {year} {2013})}\BibitemShut {NoStop}%
\bibitem [{\citenamefont {Taylor}\ \emph {et~al.}(2014)\citenamefont {Taylor}, \citenamefont {Yeung},\ and\ \citenamefont {Hancock}}]{taylor2014antibiotic}%
  \BibitemOpen
  \bibfield  {author} {\bibinfo {author} {\bibfnamefont {P.~K.}\ \bibnamefont {Taylor}}, \bibinfo {author} {\bibfnamefont {A.~T.}\ \bibnamefont {Yeung}}, \ and\ \bibinfo {author} {\bibfnamefont {R.~E.}\ \bibnamefont {Hancock}},\ }\href@noop {} {\bibfield  {journal} {\bibinfo  {journal} {Journal of biotechnology}\ }\textbf {\bibinfo {volume} {191}},\ \bibinfo {pages} {121} (\bibinfo {year} {2014})}\BibitemShut {NoStop}%
\bibitem [{\citenamefont {Morsky}\ and\ \citenamefont {Vural}(2022)}]{morsky2022suppressing}%
  \BibitemOpen
  \bibfield  {author} {\bibinfo {author} {\bibfnamefont {B.}~\bibnamefont {Morsky}}\ and\ \bibinfo {author} {\bibfnamefont {D.~C.}\ \bibnamefont {Vural}},\ }\href@noop {} {\bibfield  {journal} {\bibinfo  {journal} {Theoretical Ecology}\ }\textbf {\bibinfo {volume} {15}},\ \bibinfo {pages} {115} (\bibinfo {year} {2022})}\BibitemShut {NoStop}%
\bibitem [{\citenamefont {Zhai}\ \emph {et~al.}(2017)\citenamefont {Zhai}, \citenamefont {Wang}, \citenamefont {Shi},\ and\ \citenamefont {Long}}]{zhai2017microbial}%
  \BibitemOpen
  \bibfield  {author} {\bibinfo {author} {\bibfnamefont {J.}~\bibnamefont {Zhai}}, \bibinfo {author} {\bibfnamefont {Z.}~\bibnamefont {Wang}}, \bibinfo {author} {\bibfnamefont {P.}~\bibnamefont {Shi}}, \ and\ \bibinfo {author} {\bibfnamefont {C.}~\bibnamefont {Long}},\ }\href@noop {} {\bibfield  {journal} {\bibinfo  {journal} {PloS one}\ }\textbf {\bibinfo {volume} {12}},\ \bibinfo {pages} {e0170417} (\bibinfo {year} {2017})}\BibitemShut {NoStop}%
\bibitem [{\citenamefont {Boyle}\ \emph {et~al.}(2013)\citenamefont {Boyle}, \citenamefont {Heilmann}, \citenamefont {van Ditmarsch},\ and\ \citenamefont {Xavier}}]{boyle2013exploiting}%
  \BibitemOpen
  \bibfield  {author} {\bibinfo {author} {\bibfnamefont {K.~E.}\ \bibnamefont {Boyle}}, \bibinfo {author} {\bibfnamefont {S.}~\bibnamefont {Heilmann}}, \bibinfo {author} {\bibfnamefont {D.}~\bibnamefont {van Ditmarsch}}, \ and\ \bibinfo {author} {\bibfnamefont {J.~B.}\ \bibnamefont {Xavier}},\ }\href@noop {} {\bibfield  {journal} {\bibinfo  {journal} {Current opinion in microbiology}\ }\textbf {\bibinfo {volume} {16}},\ \bibinfo {pages} {207} (\bibinfo {year} {2013})}\BibitemShut {NoStop}%
\bibitem [{\citenamefont {Zhou}\ \emph {et~al.}(2015)\citenamefont {Zhou}, \citenamefont {Qiao}, \citenamefont {Edgar},\ and\ \citenamefont {Stephanopoulos}}]{zhou2015distributing}%
  \BibitemOpen
  \bibfield  {author} {\bibinfo {author} {\bibfnamefont {K.}~\bibnamefont {Zhou}}, \bibinfo {author} {\bibfnamefont {K.}~\bibnamefont {Qiao}}, \bibinfo {author} {\bibfnamefont {S.}~\bibnamefont {Edgar}}, \ and\ \bibinfo {author} {\bibfnamefont {G.}~\bibnamefont {Stephanopoulos}},\ }\href@noop {} {\bibfield  {journal} {\bibinfo  {journal} {Nature biotechnology}\ }\textbf {\bibinfo {volume} {33}},\ \bibinfo {pages} {377} (\bibinfo {year} {2015})}\BibitemShut {NoStop}%
\bibitem [{\citenamefont {Acar}\ \emph {et~al.}(2020)\citenamefont {Acar}, \citenamefont {Nichol}, \citenamefont {Fernandez-Mateos}, \citenamefont {Cresswell}, \citenamefont {Barozzi}, \citenamefont {Hong}, \citenamefont {Trahearn}, \citenamefont {Spiteri}, \citenamefont {Stubbs}, \citenamefont {Burke} \emph {et~al.}}]{acar2020exploiting}%
  \BibitemOpen
  \bibfield  {author} {\bibinfo {author} {\bibfnamefont {A.}~\bibnamefont {Acar}}, \bibinfo {author} {\bibfnamefont {D.}~\bibnamefont {Nichol}}, \bibinfo {author} {\bibfnamefont {J.}~\bibnamefont {Fernandez-Mateos}}, \bibinfo {author} {\bibfnamefont {G.~D.}\ \bibnamefont {Cresswell}}, \bibinfo {author} {\bibfnamefont {I.}~\bibnamefont {Barozzi}}, \bibinfo {author} {\bibfnamefont {S.~P.}\ \bibnamefont {Hong}}, \bibinfo {author} {\bibfnamefont {N.}~\bibnamefont {Trahearn}}, \bibinfo {author} {\bibfnamefont {I.}~\bibnamefont {Spiteri}}, \bibinfo {author} {\bibfnamefont {M.}~\bibnamefont {Stubbs}}, \bibinfo {author} {\bibfnamefont {R.}~\bibnamefont {Burke}},  \emph {et~al.},\ }\href@noop {} {\bibfield  {journal} {\bibinfo  {journal} {Nature communications}\ }\textbf {\bibinfo {volume} {11}},\ \bibinfo {pages} {1} (\bibinfo {year} {2020})}\BibitemShut {NoStop}%
\bibitem [{\citenamefont {Gatenby}\ \emph {et~al.}(2009)\citenamefont {Gatenby}, \citenamefont {Silva}, \citenamefont {Gillies},\ and\ \citenamefont {Frieden}}]{gatenby2009adaptive}%
  \BibitemOpen
  \bibfield  {author} {\bibinfo {author} {\bibfnamefont {R.~A.}\ \bibnamefont {Gatenby}}, \bibinfo {author} {\bibfnamefont {A.~S.}\ \bibnamefont {Silva}}, \bibinfo {author} {\bibfnamefont {R.~J.}\ \bibnamefont {Gillies}}, \ and\ \bibinfo {author} {\bibfnamefont {B.~R.}\ \bibnamefont {Frieden}},\ }\href@noop {} {\bibfield  {journal} {\bibinfo  {journal} {Cancer research}\ }\textbf {\bibinfo {volume} {69}},\ \bibinfo {pages} {4894} (\bibinfo {year} {2009})}\BibitemShut {NoStop}%
\bibitem [{\citenamefont {Enriquez-Navas}\ \emph {et~al.}(2016)\citenamefont {Enriquez-Navas}, \citenamefont {Kam}, \citenamefont {Das}, \citenamefont {Hassan}, \citenamefont {Silva}, \citenamefont {Foroutan}, \citenamefont {Ruiz}, \citenamefont {Martinez}, \citenamefont {Minton}, \citenamefont {Gillies} \emph {et~al.}}]{enriquez2016exploiting}%
  \BibitemOpen
  \bibfield  {author} {\bibinfo {author} {\bibfnamefont {P.~M.}\ \bibnamefont {Enriquez-Navas}}, \bibinfo {author} {\bibfnamefont {Y.}~\bibnamefont {Kam}}, \bibinfo {author} {\bibfnamefont {T.}~\bibnamefont {Das}}, \bibinfo {author} {\bibfnamefont {S.}~\bibnamefont {Hassan}}, \bibinfo {author} {\bibfnamefont {A.}~\bibnamefont {Silva}}, \bibinfo {author} {\bibfnamefont {P.}~\bibnamefont {Foroutan}}, \bibinfo {author} {\bibfnamefont {E.}~\bibnamefont {Ruiz}}, \bibinfo {author} {\bibfnamefont {G.}~\bibnamefont {Martinez}}, \bibinfo {author} {\bibfnamefont {S.}~\bibnamefont {Minton}}, \bibinfo {author} {\bibfnamefont {R.~J.}\ \bibnamefont {Gillies}},  \emph {et~al.},\ }\href@noop {} {\bibfield  {journal} {\bibinfo  {journal} {Science translational medicine}\ }\textbf {\bibinfo {volume} {8}},\ \bibinfo {pages} {327ra24} (\bibinfo {year} {2016})}\BibitemShut {NoStop}%
\bibitem [{\citenamefont {Gallaher}\ \emph {et~al.}(2018)\citenamefont {Gallaher}, \citenamefont {Enriquez-Navas}, \citenamefont {Luddy}, \citenamefont {Gatenby},\ and\ \citenamefont {Anderson}}]{gallaher2018spatial}%
  \BibitemOpen
  \bibfield  {author} {\bibinfo {author} {\bibfnamefont {J.~A.}\ \bibnamefont {Gallaher}}, \bibinfo {author} {\bibfnamefont {P.~M.}\ \bibnamefont {Enriquez-Navas}}, \bibinfo {author} {\bibfnamefont {K.~A.}\ \bibnamefont {Luddy}}, \bibinfo {author} {\bibfnamefont {R.~A.}\ \bibnamefont {Gatenby}}, \ and\ \bibinfo {author} {\bibfnamefont {A.~R.}\ \bibnamefont {Anderson}},\ }\href@noop {} {\bibfield  {journal} {\bibinfo  {journal} {Cancer research}\ }\textbf {\bibinfo {volume} {78}},\ \bibinfo {pages} {2127} (\bibinfo {year} {2018})}\BibitemShut {NoStop}%
\bibitem [{\citenamefont {Zhang}\ \emph {et~al.}(2017)\citenamefont {Zhang}, \citenamefont {Cunningham}, \citenamefont {Brown},\ and\ \citenamefont {Gatenby}}]{zhang2017integrating}%
  \BibitemOpen
  \bibfield  {author} {\bibinfo {author} {\bibfnamefont {J.}~\bibnamefont {Zhang}}, \bibinfo {author} {\bibfnamefont {J.~J.}\ \bibnamefont {Cunningham}}, \bibinfo {author} {\bibfnamefont {J.~S.}\ \bibnamefont {Brown}}, \ and\ \bibinfo {author} {\bibfnamefont {R.~A.}\ \bibnamefont {Gatenby}},\ }\href@noop {} {\bibfield  {journal} {\bibinfo  {journal} {Nature communications}\ }\textbf {\bibinfo {volume} {8}},\ \bibinfo {pages} {1} (\bibinfo {year} {2017})}\BibitemShut {NoStop}%
\bibitem [{\citenamefont {Szab{\'o}}\ and\ \citenamefont {Fath}(2007)}]{szabo}%
  \BibitemOpen
  \bibfield  {author} {\bibinfo {author} {\bibfnamefont {G.}~\bibnamefont {Szab{\'o}}}\ and\ \bibinfo {author} {\bibfnamefont {G.}~\bibnamefont {Fath}},\ }\href@noop {} {\bibfield  {journal} {\bibinfo  {journal} {Physics reports}\ }\textbf {\bibinfo {volume} {446}},\ \bibinfo {pages} {97} (\bibinfo {year} {2007})}\BibitemShut {NoStop}%
\bibitem [{\citenamefont {Allen}\ \emph {et~al.}(2013)\citenamefont {Allen}, \citenamefont {Gore},\ and\ \citenamefont {Nowak}}]{Allen2013}%
  \BibitemOpen
  \bibfield  {author} {\bibinfo {author} {\bibfnamefont {B.}~\bibnamefont {Allen}}, \bibinfo {author} {\bibfnamefont {J.}~\bibnamefont {Gore}}, \ and\ \bibinfo {author} {\bibfnamefont {M.~A.}\ \bibnamefont {Nowak}},\ }\href@noop {} {\bibfield  {journal} {\bibinfo  {journal} {Elife}\ }\textbf {\bibinfo {volume} {2}},\ \bibinfo {pages} {e01169} (\bibinfo {year} {2013})}\BibitemShut {NoStop}%
\bibitem [{\citenamefont {Nowak}\ and\ \citenamefont {Sigmund}(2004)}]{nowak04}%
  \BibitemOpen
  \bibfield  {author} {\bibinfo {author} {\bibfnamefont {M.~A.}\ \bibnamefont {Nowak}}\ and\ \bibinfo {author} {\bibfnamefont {K.}~\bibnamefont {Sigmund}},\ }\href@noop {} {\bibfield  {journal} {\bibinfo  {journal} {Science}\ }\textbf {\bibinfo {volume} {303}},\ \bibinfo {pages} {793} (\bibinfo {year} {2004})}\BibitemShut {NoStop}%
\bibitem [{\citenamefont {Vural}\ \emph {et~al.}(2015)\citenamefont {Vural}, \citenamefont {Isakov},\ and\ \citenamefont {L.}}]{dcv2}%
  \BibitemOpen
  \bibfield  {author} {\bibinfo {author} {\bibfnamefont {D.~C.}\ \bibnamefont {Vural}}, \bibinfo {author} {\bibfnamefont {A.}~\bibnamefont {Isakov}}, \ and\ \bibinfo {author} {\bibfnamefont {M.}~\bibnamefont {L.}},\ }\href@noop {} {\bibfield  {journal} {\bibinfo  {journal} {Proceedings of the Royal Society, B}\ }\textbf {\bibinfo {volume} {12}},\ \bibinfo {pages} {20150044} (\bibinfo {year} {2015})}\BibitemShut {NoStop}%
\bibitem [{\citenamefont {Medvinsky}\ \emph {et~al.}(2002)\citenamefont {Medvinsky}, \citenamefont {Petrovskii}, \citenamefont {Tikhonova}, \citenamefont {Malchow},\ and\ \citenamefont {Li}}]{Medvinsky2002}%
  \BibitemOpen
  \bibfield  {author} {\bibinfo {author} {\bibfnamefont {A.~B.}\ \bibnamefont {Medvinsky}}, \bibinfo {author} {\bibfnamefont {S.~V.}\ \bibnamefont {Petrovskii}}, \bibinfo {author} {\bibfnamefont {I.~A.}\ \bibnamefont {Tikhonova}}, \bibinfo {author} {\bibfnamefont {H.}~\bibnamefont {Malchow}}, \ and\ \bibinfo {author} {\bibfnamefont {B.-L.}\ \bibnamefont {Li}},\ }\href@noop {} {\bibfield  {journal} {\bibinfo  {journal} {SIAM review}\ }\textbf {\bibinfo {volume} {44}},\ \bibinfo {pages} {311} (\bibinfo {year} {2002})}\BibitemShut {NoStop}%
\bibitem [{\citenamefont {Nadell}\ \emph {et~al.}(2010)\citenamefont {Nadell}, \citenamefont {Foster},\ and\ \citenamefont {Xavier}}]{Nadell2010}%
  \BibitemOpen
  \bibfield  {author} {\bibinfo {author} {\bibfnamefont {C.~D.}\ \bibnamefont {Nadell}}, \bibinfo {author} {\bibfnamefont {K.~R.}\ \bibnamefont {Foster}}, \ and\ \bibinfo {author} {\bibfnamefont {J.~B.}\ \bibnamefont {Xavier}},\ }\href@noop {} {\bibfield  {journal} {\bibinfo  {journal} {PLoS Comput Biol}\ }\textbf {\bibinfo {volume} {6}},\ \bibinfo {pages} {e1000716} (\bibinfo {year} {2010})}\BibitemShut {NoStop}%
\bibitem [{\citenamefont {Nadell}\ \emph {et~al.}(2013)\citenamefont {Nadell}, \citenamefont {Bucci}, \citenamefont {Drescher}, \citenamefont {Levin}, \citenamefont {Bassler},\ and\ \citenamefont {Xavier}}]{nadell2013}%
  \BibitemOpen
  \bibfield  {author} {\bibinfo {author} {\bibfnamefont {C.~D.}\ \bibnamefont {Nadell}}, \bibinfo {author} {\bibfnamefont {V.}~\bibnamefont {Bucci}}, \bibinfo {author} {\bibfnamefont {K.}~\bibnamefont {Drescher}}, \bibinfo {author} {\bibfnamefont {S.~A.}\ \bibnamefont {Levin}}, \bibinfo {author} {\bibfnamefont {B.~L.}\ \bibnamefont {Bassler}}, \ and\ \bibinfo {author} {\bibfnamefont {J.~B.}\ \bibnamefont {Xavier}},\ }\href@noop {} {\bibfield  {journal} {\bibinfo  {journal} {Proceedings of the Royal Society B: biological sciences}\ }\textbf {\bibinfo {volume} {280}},\ \bibinfo {pages} {20122770} (\bibinfo {year} {2013})}\BibitemShut {NoStop}%
\bibitem [{\citenamefont {Dobay}\ \emph {et~al.}(2014)\citenamefont {Dobay}, \citenamefont {Bagheri}, \citenamefont {Messina}, \citenamefont {K{\"u}mmerli},\ and\ \citenamefont {Rankin}}]{Dobay2014}%
  \BibitemOpen
  \bibfield  {author} {\bibinfo {author} {\bibfnamefont {A.}~\bibnamefont {Dobay}}, \bibinfo {author} {\bibfnamefont {H.}~\bibnamefont {Bagheri}}, \bibinfo {author} {\bibfnamefont {A.}~\bibnamefont {Messina}}, \bibinfo {author} {\bibfnamefont {R.}~\bibnamefont {K{\"u}mmerli}}, \ and\ \bibinfo {author} {\bibfnamefont {D.}~\bibnamefont {Rankin}},\ }\href@noop {} {\bibfield  {journal} {\bibinfo  {journal} {Journal of evolutionary biology}\ }\textbf {\bibinfo {volume} {27}},\ \bibinfo {pages} {1869} (\bibinfo {year} {2014})}\BibitemShut {NoStop}%
\bibitem [{\citenamefont {Driscoll}\ and\ \citenamefont {Pepper}(2010)}]{Driscoll2010}%
  \BibitemOpen
  \bibfield  {author} {\bibinfo {author} {\bibfnamefont {W.~W.}\ \bibnamefont {Driscoll}}\ and\ \bibinfo {author} {\bibfnamefont {J.~W.}\ \bibnamefont {Pepper}},\ }\href@noop {} {\bibfield  {journal} {\bibinfo  {journal} {Evolution}\ }\textbf {\bibinfo {volume} {64}},\ \bibinfo {pages} {2682} (\bibinfo {year} {2010})}\BibitemShut {NoStop}%
\bibitem [{\citenamefont {Wakano}\ \emph {et~al.}(2009)\citenamefont {Wakano}, \citenamefont {Nowak},\ and\ \citenamefont {Hauert}}]{wakano2009}%
  \BibitemOpen
  \bibfield  {author} {\bibinfo {author} {\bibfnamefont {J.~Y.}\ \bibnamefont {Wakano}}, \bibinfo {author} {\bibfnamefont {M.~A.}\ \bibnamefont {Nowak}}, \ and\ \bibinfo {author} {\bibfnamefont {C.}~\bibnamefont {Hauert}},\ }\href@noop {} {\bibfield  {journal} {\bibinfo  {journal} {Proceedings of the National Academy of Sciences}\ }\textbf {\bibinfo {volume} {106}},\ \bibinfo {pages} {7910} (\bibinfo {year} {2009})}\BibitemShut {NoStop}%
\bibitem [{\citenamefont {Hauert}\ \emph {et~al.}(2008)\citenamefont {Hauert}, \citenamefont {Wakano},\ and\ \citenamefont {Doebeli}}]{hauert2008}%
  \BibitemOpen
  \bibfield  {author} {\bibinfo {author} {\bibfnamefont {C.}~\bibnamefont {Hauert}}, \bibinfo {author} {\bibfnamefont {J.~Y.}\ \bibnamefont {Wakano}}, \ and\ \bibinfo {author} {\bibfnamefont {M.}~\bibnamefont {Doebeli}},\ }\href@noop {} {\bibfield  {journal} {\bibinfo  {journal} {Theoretical population biology}\ }\textbf {\bibinfo {volume} {73}},\ \bibinfo {pages} {257} (\bibinfo {year} {2008})}\BibitemShut {NoStop}%
\bibitem [{\citenamefont {Menon}\ and\ \citenamefont {Korolev}(2015)}]{menon2015public}%
  \BibitemOpen
  \bibfield  {author} {\bibinfo {author} {\bibfnamefont {R.}~\bibnamefont {Menon}}\ and\ \bibinfo {author} {\bibfnamefont {K.~S.}\ \bibnamefont {Korolev}},\ }\href@noop {} {\bibfield  {journal} {\bibinfo  {journal} {Physical review letters}\ }\textbf {\bibinfo {volume} {114}},\ \bibinfo {pages} {168102} (\bibinfo {year} {2015})}\BibitemShut {NoStop}%
\bibitem [{\citenamefont {Monod}(1949)}]{Monod1949}%
  \BibitemOpen
  \bibfield  {author} {\bibinfo {author} {\bibfnamefont {J.}~\bibnamefont {Monod}},\ }\href@noop {} {\bibfield  {journal} {\bibinfo  {journal} {Annual Reviews in Microbiology}\ }\textbf {\bibinfo {volume} {3}},\ \bibinfo {pages} {371} (\bibinfo {year} {1949})}\BibitemShut {NoStop}%
\bibitem [{\citenamefont {Allen}\ and\ \citenamefont {Waclaw}(2018)}]{allen2018bacterial}%
  \BibitemOpen
  \bibfield  {author} {\bibinfo {author} {\bibfnamefont {R.~J.}\ \bibnamefont {Allen}}\ and\ \bibinfo {author} {\bibfnamefont {B.}~\bibnamefont {Waclaw}},\ }\href@noop {} {\bibfield  {journal} {\bibinfo  {journal} {Reports on Progress in Physics}\ }\textbf {\bibinfo {volume} {82}},\ \bibinfo {pages} {016601} (\bibinfo {year} {2018})}\BibitemShut {NoStop}%
\bibitem [{\citenamefont {Zhao}\ \emph {et~al.}(2016)\citenamefont {Zhao}, \citenamefont {Hemann},\ and\ \citenamefont {Lauffenburger}}]{zhao2016modeling}%
  \BibitemOpen
  \bibfield  {author} {\bibinfo {author} {\bibfnamefont {B.}~\bibnamefont {Zhao}}, \bibinfo {author} {\bibfnamefont {M.~T.}\ \bibnamefont {Hemann}}, \ and\ \bibinfo {author} {\bibfnamefont {D.~A.}\ \bibnamefont {Lauffenburger}},\ }\href@noop {} {\bibfield  {journal} {\bibinfo  {journal} {Trends in cancer}\ }\textbf {\bibinfo {volume} {2}},\ \bibinfo {pages} {144} (\bibinfo {year} {2016})}\BibitemShut {NoStop}%
\bibitem [{\citenamefont {Chuang}\ \emph {et~al.}(2009)\citenamefont {Chuang}, \citenamefont {Rivoire},\ and\ \citenamefont {Leibler}}]{chuang2009simpson}%
  \BibitemOpen
  \bibfield  {author} {\bibinfo {author} {\bibfnamefont {J.~S.}\ \bibnamefont {Chuang}}, \bibinfo {author} {\bibfnamefont {O.}~\bibnamefont {Rivoire}}, \ and\ \bibinfo {author} {\bibfnamefont {S.}~\bibnamefont {Leibler}},\ }\href@noop {} {\bibfield  {journal} {\bibinfo  {journal} {Science}\ }\textbf {\bibinfo {volume} {323}},\ \bibinfo {pages} {272} (\bibinfo {year} {2009})}\BibitemShut {NoStop}%
\bibitem [{\citenamefont {Penn}\ \emph {et~al.}(2012)\citenamefont {Penn}, \citenamefont {Conibear}, \citenamefont {Watson}, \citenamefont {Kraaijeveld},\ and\ \citenamefont {Webb}}]{penn2012can}%
  \BibitemOpen
  \bibfield  {author} {\bibinfo {author} {\bibfnamefont {A.~S.}\ \bibnamefont {Penn}}, \bibinfo {author} {\bibfnamefont {T.~C.}\ \bibnamefont {Conibear}}, \bibinfo {author} {\bibfnamefont {R.~A.}\ \bibnamefont {Watson}}, \bibinfo {author} {\bibfnamefont {A.~R.}\ \bibnamefont {Kraaijeveld}}, \ and\ \bibinfo {author} {\bibfnamefont {J.~S.}\ \bibnamefont {Webb}},\ }\href@noop {} {\bibfield  {journal} {\bibinfo  {journal} {FEMS Immunology \& Medical Microbiology}\ }\textbf {\bibinfo {volume} {65}},\ \bibinfo {pages} {226} (\bibinfo {year} {2012})}\BibitemShut {NoStop}%
\bibitem [{\citenamefont {Purcell}(1977)}]{purcell1977life}%
  \BibitemOpen
  \bibfield  {author} {\bibinfo {author} {\bibfnamefont {E.~M.}\ \bibnamefont {Purcell}},\ }\href@noop {} {\bibfield  {journal} {\bibinfo  {journal} {American journal of physics}\ }\textbf {\bibinfo {volume} {45}},\ \bibinfo {pages} {3} (\bibinfo {year} {1977})}\BibitemShut {NoStop}%
\bibitem [{\citenamefont {Coyte}\ \emph {et~al.}(2017)\citenamefont {Coyte}, \citenamefont {Tabuteau}, \citenamefont {Gaffney}, \citenamefont {Foster},\ and\ \citenamefont {Durham}}]{coyte2017microbial}%
  \BibitemOpen
  \bibfield  {author} {\bibinfo {author} {\bibfnamefont {K.~Z.}\ \bibnamefont {Coyte}}, \bibinfo {author} {\bibfnamefont {H.}~\bibnamefont {Tabuteau}}, \bibinfo {author} {\bibfnamefont {E.~A.}\ \bibnamefont {Gaffney}}, \bibinfo {author} {\bibfnamefont {K.~R.}\ \bibnamefont {Foster}}, \ and\ \bibinfo {author} {\bibfnamefont {W.~M.}\ \bibnamefont {Durham}},\ }\href@noop {} {\bibfield  {journal} {\bibinfo  {journal} {Proceedings of the National Academy of Sciences}\ }\textbf {\bibinfo {volume} {114}},\ \bibinfo {pages} {E161} (\bibinfo {year} {2017})}\BibitemShut {NoStop}%
\bibitem [{\citenamefont {Dawson}\ \emph {et~al.}(2018)\citenamefont {Dawson}, \citenamefont {Bailly}, \citenamefont {Dos~Santos}, \citenamefont {Moreno}, \citenamefont {Devilliers}, \citenamefont {Maroni}, \citenamefont {Sueur}, \citenamefont {Casali}, \citenamefont {Ujvari}, \citenamefont {Thomas} \emph {et~al.}}]{dawson2018social}%
  \BibitemOpen
  \bibfield  {author} {\bibinfo {author} {\bibfnamefont {E.~H.}\ \bibnamefont {Dawson}}, \bibinfo {author} {\bibfnamefont {T.~P.}\ \bibnamefont {Bailly}}, \bibinfo {author} {\bibfnamefont {J.}~\bibnamefont {Dos~Santos}}, \bibinfo {author} {\bibfnamefont {C.}~\bibnamefont {Moreno}}, \bibinfo {author} {\bibfnamefont {M.}~\bibnamefont {Devilliers}}, \bibinfo {author} {\bibfnamefont {B.}~\bibnamefont {Maroni}}, \bibinfo {author} {\bibfnamefont {C.}~\bibnamefont {Sueur}}, \bibinfo {author} {\bibfnamefont {A.}~\bibnamefont {Casali}}, \bibinfo {author} {\bibfnamefont {B.}~\bibnamefont {Ujvari}}, \bibinfo {author} {\bibfnamefont {F.}~\bibnamefont {Thomas}},  \emph {et~al.},\ }\href@noop {} {\bibfield  {journal} {\bibinfo  {journal} {Nature communications}\ }\textbf {\bibinfo {volume} {9}},\ \bibinfo {pages} {1} (\bibinfo {year} {2018})}\BibitemShut {NoStop}%
\bibitem [{\citenamefont {Benzekry}\ \emph {et~al.}(2014)\citenamefont {Benzekry}, \citenamefont {Lamont}, \citenamefont {Beheshti}, \citenamefont {Tracz}, \citenamefont {Ebos}, \citenamefont {Hlatky},\ and\ \citenamefont {Hahnfeldt}}]{benzekry2014classical}%
  \BibitemOpen
  \bibfield  {author} {\bibinfo {author} {\bibfnamefont {S.}~\bibnamefont {Benzekry}}, \bibinfo {author} {\bibfnamefont {C.}~\bibnamefont {Lamont}}, \bibinfo {author} {\bibfnamefont {A.}~\bibnamefont {Beheshti}}, \bibinfo {author} {\bibfnamefont {A.}~\bibnamefont {Tracz}}, \bibinfo {author} {\bibfnamefont {J.~M.}\ \bibnamefont {Ebos}}, \bibinfo {author} {\bibfnamefont {L.}~\bibnamefont {Hlatky}}, \ and\ \bibinfo {author} {\bibfnamefont {P.}~\bibnamefont {Hahnfeldt}},\ }\href@noop {} {\bibfield  {journal} {\bibinfo  {journal} {PLoS Comput Biol}\ }\textbf {\bibinfo {volume} {10}},\ \bibinfo {pages} {e1003800} (\bibinfo {year} {2014})}\BibitemShut {NoStop}%
\bibitem [{\citenamefont {Koziol}\ \emph {et~al.}(2020)\citenamefont {Koziol}, \citenamefont {Falls},\ and\ \citenamefont {Schnitzer}}]{koziol2020different}%
  \BibitemOpen
  \bibfield  {author} {\bibinfo {author} {\bibfnamefont {J.~A.}\ \bibnamefont {Koziol}}, \bibinfo {author} {\bibfnamefont {T.~J.}\ \bibnamefont {Falls}}, \ and\ \bibinfo {author} {\bibfnamefont {J.~E.}\ \bibnamefont {Schnitzer}},\ }\href@noop {} {\bibfield  {journal} {\bibinfo  {journal} {BMC cancer}\ }\textbf {\bibinfo {volume} {20}},\ \bibinfo {pages} {1} (\bibinfo {year} {2020})}\BibitemShut {NoStop}%
\bibitem [{\citenamefont {Roussos}\ \emph {et~al.}(2011)\citenamefont {Roussos}, \citenamefont {Condeelis},\ and\ \citenamefont {Patsialou}}]{roussos2011chemotaxis}%
  \BibitemOpen
  \bibfield  {author} {\bibinfo {author} {\bibfnamefont {E.~T.}\ \bibnamefont {Roussos}}, \bibinfo {author} {\bibfnamefont {J.~S.}\ \bibnamefont {Condeelis}}, \ and\ \bibinfo {author} {\bibfnamefont {A.}~\bibnamefont {Patsialou}},\ }\href@noop {} {\bibfield  {journal} {\bibinfo  {journal} {Nature Reviews Cancer}\ }\textbf {\bibinfo {volume} {11}},\ \bibinfo {pages} {573} (\bibinfo {year} {2011})}\BibitemShut {NoStop}%
\bibitem [{\citenamefont {Cheung}\ and\ \citenamefont {Ewald}(2016)}]{cheung2016collective}%
  \BibitemOpen
  \bibfield  {author} {\bibinfo {author} {\bibfnamefont {K.~J.}\ \bibnamefont {Cheung}}\ and\ \bibinfo {author} {\bibfnamefont {A.~J.}\ \bibnamefont {Ewald}},\ }\href@noop {} {\bibfield  {journal} {\bibinfo  {journal} {Science}\ }\textbf {\bibinfo {volume} {352}},\ \bibinfo {pages} {167} (\bibinfo {year} {2016})}\BibitemShut {NoStop}%
\bibitem [{\citenamefont {Cheung}\ \emph {et~al.}(2016)\citenamefont {Cheung}, \citenamefont {Padmanaban}, \citenamefont {Silvestri}, \citenamefont {Schipper}, \citenamefont {Cohen}, \citenamefont {Fairchild}, \citenamefont {Gorin}, \citenamefont {Verdone}, \citenamefont {Pienta}, \citenamefont {Bader} \emph {et~al.}}]{cheung2016polyclonal}%
  \BibitemOpen
  \bibfield  {author} {\bibinfo {author} {\bibfnamefont {K.~J.}\ \bibnamefont {Cheung}}, \bibinfo {author} {\bibfnamefont {V.}~\bibnamefont {Padmanaban}}, \bibinfo {author} {\bibfnamefont {V.}~\bibnamefont {Silvestri}}, \bibinfo {author} {\bibfnamefont {K.}~\bibnamefont {Schipper}}, \bibinfo {author} {\bibfnamefont {J.~D.}\ \bibnamefont {Cohen}}, \bibinfo {author} {\bibfnamefont {A.~N.}\ \bibnamefont {Fairchild}}, \bibinfo {author} {\bibfnamefont {M.~A.}\ \bibnamefont {Gorin}}, \bibinfo {author} {\bibfnamefont {J.~E.}\ \bibnamefont {Verdone}}, \bibinfo {author} {\bibfnamefont {K.~J.}\ \bibnamefont {Pienta}}, \bibinfo {author} {\bibfnamefont {J.~S.}\ \bibnamefont {Bader}},  \emph {et~al.},\ }\href@noop {} {\bibfield  {journal} {\bibinfo  {journal} {Proceedings of the National Academy of Sciences}\ }\textbf {\bibinfo {volume} {113}},\ \bibinfo {pages} {E854} (\bibinfo {year} {2016})}\BibitemShut {NoStop}%
\bibitem [{\citenamefont {Graner}\ and\ \citenamefont {Glazier}(1992)}]{graner1992simulation}%
  \BibitemOpen
  \bibfield  {author} {\bibinfo {author} {\bibfnamefont {F.}~\bibnamefont {Graner}}\ and\ \bibinfo {author} {\bibfnamefont {J.~A.}\ \bibnamefont {Glazier}},\ }\href@noop {} {\bibfield  {journal} {\bibinfo  {journal} {Physical review letters}\ }\textbf {\bibinfo {volume} {69}},\ \bibinfo {pages} {2013} (\bibinfo {year} {1992})}\BibitemShut {NoStop}%
\bibitem [{\citenamefont {Szab{\'o}}\ and\ \citenamefont {Merks}(2013)}]{szabo2013cellular}%
  \BibitemOpen
  \bibfield  {author} {\bibinfo {author} {\bibfnamefont {A.}~\bibnamefont {Szab{\'o}}}\ and\ \bibinfo {author} {\bibfnamefont {R.~M.}\ \bibnamefont {Merks}},\ }\href@noop {} {\bibfield  {journal} {\bibinfo  {journal} {Frontiers in oncology}\ }\textbf {\bibinfo {volume} {3}},\ \bibinfo {pages} {87} (\bibinfo {year} {2013})}\BibitemShut {NoStop}%
\bibitem [{\citenamefont {Barton}\ \emph {et~al.}(2017)\citenamefont {Barton}, \citenamefont {Henkes}, \citenamefont {Weijer},\ and\ \citenamefont {Sknepnek}}]{barton2017active}%
  \BibitemOpen
  \bibfield  {author} {\bibinfo {author} {\bibfnamefont {D.~L.}\ \bibnamefont {Barton}}, \bibinfo {author} {\bibfnamefont {S.}~\bibnamefont {Henkes}}, \bibinfo {author} {\bibfnamefont {C.~J.}\ \bibnamefont {Weijer}}, \ and\ \bibinfo {author} {\bibfnamefont {R.}~\bibnamefont {Sknepnek}},\ }\href@noop {} {\bibfield  {journal} {\bibinfo  {journal} {PLoS computational biology}\ }\textbf {\bibinfo {volume} {13}},\ \bibinfo {pages} {e1005569} (\bibinfo {year} {2017})}\BibitemShut {NoStop}%
\bibitem [{\citenamefont {Koride}\ \emph {et~al.}(2018)\citenamefont {Koride}, \citenamefont {Loza},\ and\ \citenamefont {Sun}}]{koride2018epithelial}%
  \BibitemOpen
  \bibfield  {author} {\bibinfo {author} {\bibfnamefont {S.}~\bibnamefont {Koride}}, \bibinfo {author} {\bibfnamefont {A.~J.}\ \bibnamefont {Loza}}, \ and\ \bibinfo {author} {\bibfnamefont {S.~X.}\ \bibnamefont {Sun}},\ }\href@noop {} {\bibfield  {journal} {\bibinfo  {journal} {APL bioengineering}\ }\textbf {\bibinfo {volume} {2}},\ \bibinfo {pages} {031906} (\bibinfo {year} {2018})}\BibitemShut {NoStop}%
\bibitem [{\citenamefont {Bangerth}\ \emph {et~al.}(2007)\citenamefont {Bangerth}, \citenamefont {Hartmann},\ and\ \citenamefont {Kanschat}}]{bangerth2007deal}%
  \BibitemOpen
  \bibfield  {author} {\bibinfo {author} {\bibfnamefont {W.}~\bibnamefont {Bangerth}}, \bibinfo {author} {\bibfnamefont {R.}~\bibnamefont {Hartmann}}, \ and\ \bibinfo {author} {\bibfnamefont {G.}~\bibnamefont {Kanschat}},\ }\href@noop {} {\bibfield  {journal} {\bibinfo  {journal} {ACM Transactions on Mathematical Software (TOMS)}\ }\textbf {\bibinfo {volume} {33}},\ \bibinfo {pages} {24} (\bibinfo {year} {2007})}\BibitemShut {NoStop}%
\bibitem [{\citenamefont {Kim}(1996)}]{Kim1996}%
  \BibitemOpen
  \bibfield  {author} {\bibinfo {author} {\bibfnamefont {Y.-C.}\ \bibnamefont {Kim}},\ }\href@noop {} {\bibfield  {journal} {\bibinfo  {journal} {Korean Journal of Chemical Engineering}\ }\textbf {\bibinfo {volume} {13}},\ \bibinfo {pages} {282} (\bibinfo {year} {1996})}\BibitemShut {NoStop}%
\bibitem [{\citenamefont {Ma}\ \emph {et~al.}(2005)\citenamefont {Ma}, \citenamefont {Zhu}, \citenamefont {Ma},\ and\ \citenamefont {Yu}}]{Ma2005}%
  \BibitemOpen
  \bibfield  {author} {\bibinfo {author} {\bibfnamefont {Y.}~\bibnamefont {Ma}}, \bibinfo {author} {\bibfnamefont {C.}~\bibnamefont {Zhu}}, \bibinfo {author} {\bibfnamefont {P.}~\bibnamefont {Ma}}, \ and\ \bibinfo {author} {\bibfnamefont {K.}~\bibnamefont {Yu}},\ }\href@noop {} {\bibfield  {journal} {\bibinfo  {journal} {Journal of Chemical \& Engineering Data}\ }\textbf {\bibinfo {volume} {50}},\ \bibinfo {pages} {1192} (\bibinfo {year} {2005})}\BibitemShut {NoStop}%
\bibitem [{\citenamefont {Rusconi}\ and\ \citenamefont {Stocker}(2015)}]{rusconi2015microbes}%
  \BibitemOpen
  \bibfield  {author} {\bibinfo {author} {\bibfnamefont {R.}~\bibnamefont {Rusconi}}\ and\ \bibinfo {author} {\bibfnamefont {R.}~\bibnamefont {Stocker}},\ }\href@noop {} {\bibfield  {journal} {\bibinfo  {journal} {Current opinion in microbiology}\ }\textbf {\bibinfo {volume} {25}},\ \bibinfo {pages} {1} (\bibinfo {year} {2015})}\BibitemShut {NoStop}%
\bibitem [{\citenamefont {Busscher}\ and\ \citenamefont {van~der Mei}(2006)}]{busscher2006microbial}%
  \BibitemOpen
  \bibfield  {author} {\bibinfo {author} {\bibfnamefont {H.~J.}\ \bibnamefont {Busscher}}\ and\ \bibinfo {author} {\bibfnamefont {H.~C.}\ \bibnamefont {van~der Mei}},\ }\href@noop {} {\bibfield  {journal} {\bibinfo  {journal} {Clinical microbiology reviews}\ }\textbf {\bibinfo {volume} {19}},\ \bibinfo {pages} {127} (\bibinfo {year} {2006})}\BibitemShut {NoStop}%
\bibitem [{\citenamefont {Gibson}\ \emph {et~al.}(2018)\citenamefont {Gibson}, \citenamefont {Wilson}, \citenamefont {Feil},\ and\ \citenamefont {Eyre-Walker}}]{gibson2018distribution}%
  \BibitemOpen
  \bibfield  {author} {\bibinfo {author} {\bibfnamefont {B.}~\bibnamefont {Gibson}}, \bibinfo {author} {\bibfnamefont {D.~J.}\ \bibnamefont {Wilson}}, \bibinfo {author} {\bibfnamefont {E.}~\bibnamefont {Feil}}, \ and\ \bibinfo {author} {\bibfnamefont {A.}~\bibnamefont {Eyre-Walker}},\ }\href@noop {} {\bibfield  {journal} {\bibinfo  {journal} {Proceedings of the Royal Society B: Biological Sciences}\ }\textbf {\bibinfo {volume} {285}},\ \bibinfo {pages} {20180789} (\bibinfo {year} {2018})}\BibitemShut {NoStop}%
\bibitem [{\citenamefont {Drake}\ \emph {et~al.}(1998)\citenamefont {Drake}, \citenamefont {Charlesworth}, \citenamefont {Charlesworth},\ and\ \citenamefont {Crow}}]{drake1998rates}%
  \BibitemOpen
  \bibfield  {author} {\bibinfo {author} {\bibfnamefont {J.~W.}\ \bibnamefont {Drake}}, \bibinfo {author} {\bibfnamefont {B.}~\bibnamefont {Charlesworth}}, \bibinfo {author} {\bibfnamefont {D.}~\bibnamefont {Charlesworth}}, \ and\ \bibinfo {author} {\bibfnamefont {J.~F.}\ \bibnamefont {Crow}},\ }\href@noop {} {\bibfield  {journal} {\bibinfo  {journal} {Genetics}\ }\textbf {\bibinfo {volume} {148}},\ \bibinfo {pages} {1667} (\bibinfo {year} {1998})}\BibitemShut {NoStop}%
\end{thebibliography}%

\appendix
\section{Existence and stability conditions}
\label{app:turing}
\subsection{Steady states}
To gain an analytical understanding, we find what controls lead to extinct states versus stable states and perform a Turing analysis to see what controls lead to pattern formation.

We start by computing the homogeneous steady states $n^*, c_1^*, c_2^*$ by setting spatial and time derivative terms to zero. We then have,
\begin{align}
    n^* \left[ \alpha_1 \frac{c_1^*}{c_1^*+k_1} - \alpha_2 \frac{c_2^*}{c_2^* + k_2} - \beta_1 s_1 \right] = 0 \\
    -\lambda_1 c_1^* + n s_1 + \mu_1 = 0 \\
    -\lambda_2 c_2^* + n s_2 + \mu_2 = 0
\end{align}
Solving for $n^*, c_1^*, c_2^*$ we get for the chemicals,
\begin{align}
    c_1^* = \frac{n^* s_1 + \mu_1}{\lambda_1} \\
    c_2^* = \frac{n^* s_2 + \mu_2}{\lambda_2}
\end{align}
The microbe population steady state $n^*$ is given by solving a quadratic equation $a n^{*2} + b n^* + c = 0$ with coefficients given as,
\begin{align*}
    a = s_1 s_2 ( \alpha_1 - \alpha_2 - \beta_1 s_1) \\
    b = (s_2 \mu_1 + s_1 \mu_2)(\alpha_1 - \alpha_2 - \beta_1 s_1) \nonumber \\ - k_1 s_2 \lambda_1 (\alpha_2 + \beta_1 s_1) + k_2 s_1 \lambda_2 ( \alpha_1 - \beta_1 s_1) \\
   c =  k_2 (\alpha_1 - s_1 \beta_1) \lambda_2 \mu_1 + (\alpha_1 - \alpha_2 - \beta_1 s_1) \mu_1 \mu_2 \nonumber \\ - k_1 \lambda_1 ( \alpha_2 \mu_2 + \beta_1 s_1 (k_2 \lambda_2 + \mu_2 ) )
\end{align*}
Descartes' rule of signs states that the number of positive roots of a polynomial is at most the number of sign changes in the coefficients, not including zero coefficients, and that the difference between the number of sign changes and number of roots is even. In particular, this implies zero or one sign changes corresponds to exactly zero or one positive roots respectively. Therefore, to have a positive real solution for $n$, we need at least one sign change. Since we enforce $\alpha_1 - \alpha_2 - \beta_1 s_1 < 0$ for stability, the coefficient $a < 0$. We must therefore have either $b > 0$, $c > 0$, or both. Given a set of parameters, this defines a space of values for $\mu_1$ and $\mu_2$ for which a positive solution can exist. 

\subsection{Linear stability}
We next explore the conditions for the steady states to be linearly stable. We start by linearizing our system of equations. Our linearized system is given as
\[
\frac{\partial \mathbf{w}}{\partial t} = A \mathbf{w}
\]
where $\mathbf{w} = (n, c_1, c_2)^T - (n^*, c_1^*, c_2^*)^T$ is a perturbation from the steady state and the linear stability matrix $A$ is given as
\[ 
A = \begin{pmatrix}
0   & f_1 & f_2 \\
s_1 & -\lambda_1 & 0 \\
s_2 & 0 & - \lambda_2
\end{pmatrix}
\]
Here $f_1 = \frac{\alpha_1 k_1 n^*}{(c_1^* + k_1)^2}$ and $f_2 = - \frac{\alpha_2 k_2 n^*}{(c_2^* + k_2)^2} $. The steady state is stable if the eigenvalues of the stability matrix $A$ all have negative real part. The characteristic polynomial for $A$ is given in terms of the invariants of a $3 \times 3$ matrix,
\[
\Lambda^3 -\mathrm{tr}(A) \Lambda^2 + \frac{1}{2} \left[ \left(\mathrm{tr}(A)\right)^2 - \mathrm{tr}(A^2) \right] \Lambda - \mathrm{det}(A) = 0,
\]
for eigenvalues $\Lambda$. Using Descartes' rule of signs again, we see that the condition for all roots of the polynomial to be negative is if there are no sign changes. Since we already have $\mathrm{tr}(A) < 0$, the conditions for stability are
\begin{align}
\label{eq:stability_conditions}
\mathrm{det}(A) < 0 \nonumber \\
\left(\mathrm{tr}(A)\right)^2 - \mathrm{tr}(A^2) > 0 .
\end{align}
These conditions also give a space of valid control functions $\mu_1$ and $\mu_2$ to allow for a non-zero steady state. 

\subsection{Turing instability}
Next we include diffusion and explore the conditions giving rise to a Turing instabiilty. With diffusion, our linearized system is now,
\[
\frac{\partial \mathbf{w}}{\partial t} = D \nabla^2 \mathbf{w} + A \mathbf{w},
\]
where
\[
D = \begin{pmatrix} d_b & 0 & 0 \\
0 & d_1 & 0 \\
0 & 0 & d_2 \\
\end{pmatrix}
\]
is the diffusion matrix. To investigate the unstable wavelengths, we expand our solution in terms of Fourier modes as,
\[
\mathbf{w} (\mathbf{x}, t) = \sum_k c_k e^{i \mathbf{k} \cdot \mathbf{x}} e^{\Lambda{k} t} .
\]
Plugging this into our linearized system, we get the eigenvalue equation $(-k^2 D + A) \mathbf{w} = \Lambda \mathbf{w}$, giving a dispersion relation for eigenvalues $\Lambda(\mathbf{k})$. From this relation, we can determine which wave modes are unstable as those that correspond to positive eigenvalues $\Lambda(\mathbf{k}) > 0$. These modes are unstable in the linearized system and will grow until saturated by the nonlinear terms of our system, forming spatial patterns.

If we denote by $M(k) = -k^2 D + A$, the characteristic equation for this system is now given as

\begin{align*}
\Lambda(k)^3 - \mathrm{tr}(M(k)) \Lambda^2 + \frac{1}{2} \big{[} (\mathrm{tr}(M(k)))^2 - \mathrm{tr}(M(k)^2) \big{]} \Lambda(k) \\
- \mathrm{det}(M(k)) = 0
\end{align*}
We now use Descartes' rule of signs once again, this time to get an instability. First, we have $\mathrm{tr}(M) = -k^2 \mathrm{tr}(D) + \mathrm{tr}(A) < 0$, which makes the quadratic term positive. For the linear term, we have
\begin{align*}
(\mathrm{tr} (M(k)))^2 - \mathrm{tr}(M(k)^2) = \\
 [ -k^2 \mathrm{tr}(D) + \mathrm{tr}(A)]^2 
- \mathrm{tr}([-k^2 D + A]^2) =\\
  k^4 (\mathrm{tr}(D))^2 - 2 k^2 \mathrm{tr}(D) \mathrm{tr}(A) + (\mathrm{tr}(A))^2 \\ - \left[ k^4 \mathrm{tr}(D^2) - k^2 \mathrm{tr}(DA) -k^2 \mathrm{tr}(AD) + \mathrm{tr}(A)^2 \right] = \\
 k^4 [(\mathrm{tr}(D))^2 - \mathrm{tr}(D^2)] + (\mathrm{tr}(A))^2 - \mathrm{tr}(A^2)  \\ + k^2[ \mathrm{tr}(DA) + \mathrm{tr}(AD) - 2 \mathrm{tr}(D) \mathrm{tr}(A)] .
\end{align*}
Now since $(\mathrm{tr}(D))^2 - \mathrm{tr}(D^2) = (d_b + d + d_w)^2 -  (d_b^2 + d^2 + d_w^2) > 0$ and \ $\mathrm{tr}(DA) + \mathrm{tr}(AD) - 2 \mathrm{tr}(D) \mathrm{tr}(A) = 2(d_w \lambda +  d \lambda_w + d_b ( \lambda + \lambda_w )) > 0$, and since we require $(\mathrm{tr}(A))^2 - \mathrm{tr}(A^2) > 0$ from linear stability, the linear term is also positive. Therefore, by Descartes rule of signs, the requirement for Turing instability reduces to,

\begin{align*}
\mathrm{det}(-k^2 D + A) > 0 ,
\end{align*}
for some range of $k > 0$. 




\section{Model implementation details}
\label{app:model}
\subsection{Numerical methods}
Here we provide additional details on the implementation of our agent based stochastic simulations with chemical fields implemented via a finite element method in two spatial dimensions. Our simulation code is written in C++ using the open source deal.II library for implementing the finite element fields \cite{bangerth2007deal}. The code is available on Github: \url{https://github.com/garyuppal/MicrobeSimulator.git}. 

Our algorithm evolves microbial agents and chemical fields at each time step $\Delta t$. Microbes diffuse and advect via a biased random walk scheme. At each time step, microbes move an amount $\delta = \sqrt{4 d_b \Delta t}$ in a direction up, down, left, or right with equal probability. A bias given by $\vec{v} \Delta t$ is added for the advection term. We note that $\Delta t$ is sufficiently small for this to give an accurate representation of the diffusion-advection process with our given diffusion and flow parameters. 

Microbes also secrete chemicals locally onto a finite element mesh that then diffuse and advect using a finite element scheme. We use a backward Euler method for the time-stepping and treat diffusion, advection, and decay terms implicitly, and the source and control terms explicitly. Specifically, we discretize in time as,
\begin{align*} 
\frac{c^n(\mathbf{x}) - c^{n-1}(\mathbf{x})}{\Delta t} = d \nabla^2 c^n(\mathbf{x}) - \mathbf{v} \cdot \nabla c^n(\mathbf{x})  \\ - \lambda c^n(\mathbf{x}) +  \sum_b s_b \delta(\mathbf{x}-\mathbf{x}^{n-1}_b) + \mu^{n-1}
\end{align*}
where $c^n(\mathbf{x})$ is either chemical field at time step $n$, the sum is over all microbes at the current time step, $s_b$ is the secretion rate of microbe $b$, $\mathbf{x}_b^{n-1}$ is the position of microbe $b$ at time step $n-1$ and $\Delta t$ is our time step taken to be $\Delta t = 0.001$.

Due to the advection term, the numerical implementation of this system is not stable in general, and we need to add some sort of stabilization scheme. This can be done using discontinuous elements, using an upwinding scheme, or in a variety of other ways. We introduce an additional artificial viscosity term $\nu(c(\mathbf{x}))$ to stabilize the advection term. 

The weak form of our system is then given by,
\begin{align*} 
[1 + \lambda \Delta t](\phi, c^n ) + d  \Delta t ( \nabla \phi, \nabla  c^n ) +  \Delta t ( \nabla \phi, \nu(\mathbf{x}) \nabla  c^n ) \\
+ \Delta t (\phi,  \mathbf{v} \cdot \nabla c^n ) =  ( \phi, c^{n-1} )  +  (\phi, \mu^{n-1} ) \\
+  \sum_b s_b (\phi,  \delta(\mathbf{x}-\mathbf{x}^{n-1}_b) ) \quad \forall \phi
\end{align*}
where we use the common notation $(a,b) = \int_\Omega a b$ for the integral over the domain $\Omega$ and $\phi$ is a test function.

We discretize in space with a finite element mesh using first order Lagrange elements on quadrilateral cells in 2-dimensions (Q1 elements). Our space is then divided into cells $K$ and the integrals are approximated as $\int_\Omega \approx \sum_K \int_K$. Our stabilzation term is then given on each cell as,
\[
\nu|_K = \beta ||\mathbf{v}||_{L^\infty(K)} h_K
\]
where $h_K$ is the diameter of cell $K$, $||\cdot ||_{L^\infty(K)} $ is the norm on cell $K$, and $\beta$ is a stabilization constant chosen from initial numerical experiments and taken to be $\beta = 0.1$.

After discretizing and expanding our solution in the basis of our test functions (Galerkin method), our final scheme then amounts to solving the linear system
\[ S_{ij} U_j = F_i \]
for the solution vector $U_j$ at each time step, where $i,j$ run over all degrees of freedom and the system matrix $S_{ij}$ is given by
\begin{align*}
    S_{ij} = \sum_K  \int_K (1 + \lambda \Delta t ) \phi_i \phi_j + \Delta t \int_K   (d + \nu) \nabla \phi_i \nabla \phi_j \\
    + \Delta t \int_K \phi_i  (\mathbf{v} \cdot \nabla \phi_j)
\end{align*}
The right hand side term is given as
\[
F_i =   \sum_K  \sum_b s_b \int_K \phi_i  \delta(\mathbf{x}-\mathbf{x}^{n-1}_b) + \int_K \phi_i \mu^{n-1} .
\]

For the flow profiles in Fig. \ref{fig:shear}, we used analytical expressions for the function $\mathbf{v}(\mathbf{x})$. For our geometric control strategies however (Fig. \ref{fig:geometrycontrol}), we numerically calculate the flow field $\mathbf{v}(\mathbf{x})$ by solving the Stokes equations given as,
\begin{align*}
    -2 \mathrm{div} \ \varepsilon(\mathbf{v}) + \nabla p = \mathbf{f} \\
    - \mathrm{div} \ \mathbf{v} = 0
\end{align*}
where $\varepsilon(\mathbf{v}) = \frac{1}{2} \left[(\nabla \mathbf{v}) + (\nabla \mathbf{v})^T \right]$ is the rank-2 tensor of symmetrized gradients of the velocity field. We take external forces to be zero $\mathbf{f} = 0$. For boundary conditions, we set the velocity at the inlet (left wall) to be a constant given by an input parameter, do not impose boundary conditions on the outlet (right wall), and impose no slip boundary conditions at all other boundaries.



% ...

In the case where the flow is given from the Stokes solution, we first solve for $\mathbf{v}$ using a finite element scheme and adaptively refine our mesh in regions where the pressure gradient is largest. We then use this solution in constructing the system matrix for the finite element system for the chemical equations.

\subsection{Parameter selection and sensitivity}
Most of our parameters are taken from empirical sources as cited in Table \ref{tab:parameters}. We discuss here the relevance of these values to empirical data and some of the effects of varying these parameters.

Our model contains parameters for growth rates (public good benefit and waste harm), secretion cost, decay rates, secretion rates, diffusion constants, and mutation rates. Chemical concentrations values can be rescaled to remove saturation constants. Saturation constants therefore do not play a major role.

We first note that fitness constant values are constrained by the relations in our Turing analysis (Appendix \ref{app:turing}). In particular, we set $\alpha_1 < \alpha_2$ and $\alpha_1 > \beta s_1$ to ensure we have positive and stable steady state values for the microbial population. We fix the public good decay rate to $\lambda_1 = 50$ throughout and set the microbe diffusion constant to $d_b = 0.4$ for all but the filter where $d_b = 0.2$. We can take these to correspond to $\lambda_1 = 50 \times 10^{-4} \ \mathrm{s}^{-1}$ and $d_b = 0.4 \times 10^{-6} \ \mathrm{cm}^2 \mathrm{s}^{-1}$, corresponding to time and space units $[t] = 10^4 \ \mathrm{s}$ and $[x] = 0.1 \ \mathrm{cm}$, respectively. The largest growth rates in our model is then typically around a division every $10^{-3} \ \mathrm{s}^{-1}$ which is around empirical values of once every 20 to 30 minutes at most \cite{gibson2018distribution}. 

Diffusion constants correspond to empirical values typically on the order of $10^{-6} \mathrm{cm}^2 \mathrm{s}^{-1}$ and can vary over a couple orders of magnitude \cite{Kim1996, Ma2005}. We explore various diffusion regimes in our previous study \cite{uppal2018}. Here we first fix diffusion constants to give rise to Turing patterns and then explore the effects of various control schemes. 

We varied chemical control amplitudes over a representative range for patterns (Fig. \ref{fig:constantcontrol}a) and social behavior (Fig. \ref{fig:constantcontrol}c), starting with a spot forming state. We then chose control amplitudes that correspond to largely influencing the groups without causing extinction and varied the temporal duration of each control chemical in Fig. \ref{fig:pulsecontrol}.

Filter parameters were chosen to limit channel death and natural group reproduction. For this, we chose larger waste diffusion so that cooperating groups in a channel do not over-pollute themselves. We also chose lower fitness constants and lower microbial diffusion, to slow down the natural group reproduction rate and to have the channels be the main cause of group fragmentation. This helps to not complicate results with large natural group fragmentation rates. 

Mixer parameters were chosen to allow stable semi-cheaters and for groups to come into contact without killing each other off. In particular, we increased the waste decay rate so groups can merge without over-polluting each other as they combine. We also increased the waste diffusion so groups do not die out in the narrow channel portions of the funnel.

Mutation and flow rates were also chosen from empirical studies \cite{drake1998rates,rusconi2015microbes}. Flow rates were varied to give different shear rates in a Couette flow and were fixed in pipe and vortex geometries to cover a representative range of shear rates over the spatial domain. 

Finally, we note we investigated each type of control separately for simplicity. Combining various control strategies may lead to more robust or efficient methods for controlling population and sociality. For example, if the waste diffusion is small for groups passing through a filter, we find the population in the microchannels can die out. Adding an external source of public good might help save the population while the channels help filter out cheaters and control the sociality. We also note we do not vary much the secretion and decay rates of the chemicals. Our previous studies show the effects of varying these quantities can be qualitatively understood by rescaling the chemical concentrations (c.f. Supplementary Section of \cite{uppal2020evolution}). 

\end{document}